\documentclass[preprint,aps,prd,showpacs,floatfix,nofootinbib,superscriptaddress]{revtex4}
\usepackage{graphicx}
\usepackage{amsmath}
\usepackage{amssymb}
\usepackage{bm}
\usepackage{bbm}
\begin{document}
\title{\mbox{}\\[10pt]
Inclusive Charm Production in $\bm{\Upsilon(nS)}$ Decay
}
\author{Daekyoung~Kang}
\affiliation{Department of Physics, Korea University, Seoul 136-701, Korea}
\affiliation{Physics Department, Ohio State University, Columbus, Ohio 43210, USA}
\author{Taewon~Kim}
\affiliation{Department of Physics, Korea University, Seoul 136-701, Korea}
\author{Jungil~Lee}
\affiliation{Department of Physics, Korea University, Seoul 136-701, Korea}
\affiliation{Physics Department, Ohio State University, Columbus, Ohio 43210, USA}
\author{Chaehyun~Yu}
\affiliation{Department of Physics, Korea University, Seoul 136-701, Korea}

\begin{abstract}
Based on the NRQCD factorization formalism, we calculate the inclusive
charm production rate in $\Upsilon(nS)$ decay at 
leading order in the strong coupling constant $\alpha_s$ and the 
relative velocity $v$ of the $b$ quark in the quarkonium rest frame.
The branching fractions for $\Upsilon(nS)$ to charm for $n=1$, $2$, and $3$ 
are all around $7$--$\,9$\%. About $60$\% of the branching fraction
into charm is from annihilation of the color-singlet $b\bar{b}$ pair
into $\gamma^*\to c\bar{c}$. Most of the remaining branching fraction
is from annihilation of the color-singlet $b\bar{b}$ pair decaying into 
$c\bar{c}gg$. We also compute the momentum distributions of the charm 
quark and charmed hadrons in the decays. The virtual-photon contribution
to the charm quark momentum distribution is concentrated at the end point
while the $c\bar{c}gg$ contribution is broad. 
The momentum distributions for charmed hadrons are obtained by convolving
the charm-quark momentum distribution
with charm fragmentation functions. This makes the momentum distributions
for charmed hadrons softer than that for the charm quark.
This effect is particularly significant in the virtual-photon contribution.
\end{abstract}



\maketitle


\section{Introduction 
\label{sec:intro}}

According to the Nonrelativistic QCD (NRQCD) factorization formalism,
an annihilation decay rate of a spin-triplet $S$-wave (${}^3S_1$) 
bottomonium $\Upsilon$\footnote{
Throughout this paper, we suppress the identifier $nS$ in $\Upsilon(nS)$, 
where $n$ is the radial quantum number, unless it is necessary.}
is expressed as an infinite series of NRQCD matrix elements with 
corresponding short-distance coefficients~\cite{Bodwin:1994jh}.
The NRQCD matrix elements, which reflect the long-distance nature
of the quarkonium, scale as powers of the bottom-quark velocity $v$
in the quarkonium rest frame, which is $v^2 \approx 0.1$.
At leading order in $v$, the inclusive decay rate of the $\Upsilon$
is dominated by the color-singlet spin-triplet contribution
whose NRQCD matrix element is
$\langle \mathcal{O}_1({}^3S_1) \rangle_{\Upsilon}=%
\langle \Upsilon|\mathcal{O}_1({}^3S_1) |\Upsilon\rangle$, which is 
defined in Ref.~\cite{Bodwin:1994jh}. The subscript $1$ on the NRQCD
four-quark operator $\mathcal{O}_1$ denotes that it is a color-singlet
operator. Thus, at leading order in $v$, the inclusive light-hadronic 
decay rate of the $\Upsilon$ is expressed in a factorized form:
\begin{equation}
\Gamma[\Upsilon \to X] =
C_1\, \frac{\langle \mathcal{O}_1({}^3S_1)\rangle_{\Upsilon}}{m_b^2},
\label{Gam-ups}%
\end{equation}
where $X$ represents all possible light hadronic final states
into which $\Upsilon$ can decay and $m_b$ is the bottom-quark mass.
The short-distance coefficient $C_1$, which
is insensitive to the long-distance nature of the $\Upsilon$,
can be calculated perturbatively.
The dimensions of the matrix element 
$\langle \mathcal{O}_1({}^3S_1)\rangle_{\Upsilon}$
is $3$ so that $C_1$ is dimensionless.

At leading order in $\alpha_s$, the dominant color-singlet contribution
to $C_1$ comes from $b\bar{b}_1({}^3S_1)\to ggg$ mode, where
the three gluons are attached to the bottom-quark line.
Here, $\alpha_s$ is the strong coupling constant and $b\bar{b}_1({}^{2s+1}L_J)$
is the color-singlet $b\bar{b}$ pair with spin $s$, orbital angular 
momentum $L$, and total angular momentum $J$. 
The leading contribution
of order $\alpha_s^3$ to $C_1$ is
known through the orthopositronium decay 
rate obtained by Caswell, Lepage, and Sapirstein~\cite{Caswell:1976nx}.
The order-$\alpha_s^4$ corrections to $C_1$ were  calculated by Mackenzie
and Lepage~\cite{Mackenzie:1981sf}. This result 
was confirmed recently
by Campbell, Maltoni, and Tramontano~\cite{Campbell:2007ws}.

In addition to the three-gluon mode, $C_1$ may include the virtual-photon
contribution from $b\bar{b}_1({}^3S_1)\to\gamma^*\to q\bar{q}+X$. 
The decay rate is of order
$e_b^2e_q^2 \alpha^2$, where $\alpha$ is the
QED coupling constant 
and $e_q$ is the fractional electric
charge of the quark $q$: $e_q=\tfrac{2}{3}$ for an up-type quark and 
$-\tfrac{1}{3}$ for a down-type quark. That electromagnetic decay rate 
may appear to be highly suppressed compared to the three-gluon mode of 
order $\alpha_s^3$. However, we can make a rough estimate of the 
branching fraction $\textrm{Br}[\Upsilon\to\gamma^*\to q\bar{q}]$ 
by using the measured branching fractions for 
$\textrm{Br}[\Upsilon\to e^+e^-]$~\cite{Yao:2006px}:
$\textrm{Br}[\Upsilon\to \gamma^\ast\to q\bar{q}]\approx%
N_c\textrm{Br}[\Upsilon\to e^+e^-]\sum_q e_q^2$, where the sum is over
the four flavors of quarks lighter than the bottom
and $N_c=3$ is the number of colors.
According to this estimate, 
$\textrm{Br}[\Upsilon\to \gamma^*\to q \bar{q}]\approx 6$--$\, 8 \%$,
which may not be negligible.

At higher orders in $v$, the NRQCD factorization formula (\ref{Gam-ups})
must include additional contributions from higher Fock states which 
involve
color-octet pairs $b\bar{b}_8({}^{2s+1}L_J)$ as well as the color-singlet
ones, which are suppressed compared to the leading contribution in 
Eq.~(\ref{Gam-ups}). The order-$v^2$ and order-$v^4$ 
relativistic corrections to 
the color-singlet contributions were calculated 
by Keung and Muzinich~\cite{Keung:1982jb} and by 
Bodwin and Petrelli~\cite{Bodwin:2002hg}, respectively. 
Some of the color-octet contributions were also calculated in 
Refs.~\cite{Cheung:1996mh,Maltoni:1998nh,Brambilla:2007cz}.

Because the $\Upsilon$ is heavy enough, the decay products may include
a pair of charmed hadrons. However, unlike the light-hadronic decay mode of
the $\Upsilon$, there has been little previous work on open-charm
production in $\Upsilon$ decay. In 1978, Fritzsch and Streng predicted
the branching fraction of the decay of $\Upsilon$ into charm to be a few 
percents~\cite{Fritzsch:1978ey}, where they considered 
$\Upsilon\to ggg^*$ followed by $g^*\to c\bar{c}$. 
In 1979, Bigi and Nussinov took into account
a fusion process $\Upsilon\to c\bar{c}g$ of order
$\alpha_s^5$, in which a pair of virtual gluons create the $c\bar{c}$ 
pair~\cite{Bigi:1978tj}. In 1992, ARGUS experiment searched for charm 
production in direct decays of the $\Upsilon(1S)$ to find only an upper
limit of $\textrm{Br}^{\textrm{dir}}%
[\Upsilon(1S)\to D^*(2010)^{\pm}+X]<0.019$~\cite{Albrecht:1992ap}.

Recent runs of the CLEO III experiment have produced a large amount of
data samples at the $\Upsilon(1S)$, $\Upsilon(2S)$, and $\Upsilon(3S)$ 
resonances. The $B$-factory experiments BABAR and Belle have 
accumulated data for $\Upsilon(2S)$ and $\Upsilon(3S)$ provided by
initial-state radiation. The Belle Collaboration has also collected
data by running on the $\Upsilon(3S)$ resonance. With these 
high-luminosity data, one can now indeed study  
open-charm production in $\Upsilon$ decay. 
Very recently, some of the authors have calculated the total 
production rates and momentum distributions of the charm quark and
charmed hadrons, respectively, in the $P$-wave bottomonium decays 
$\chi_{bJ}\to c+X$ for $J=0$, 1, and 2 by using the NRQCD factorization
formalism~\cite{Bodwin:2007zf}. In order to calculate the momentum
distribution of the charmed hadrons they used the momentum distribution 
of charmed hadrons measured by the Belle Collaboration in $e^+e^-$ 
annihilation~\cite{Seuster:2005tr}.

In this work, as an extension of a previous study~\cite{Bodwin:2007zf},
we consider inclusive
charm production in the spin-triplet $S$-wave bottomonium decay.
At leading order in $v$, the dominant mechanism for the decay is
a color-singlet channel $b\bar{b}_1({}^3S_1)\to ggg^*$ 
followed by $g^*\to c\bar{c}$. As we have described earlier, 
the color-singlet mode may have significant virtual-photon 
contribution from
$b\bar{b}_1({}^3S_1)\to \gamma^*$ followed by $\gamma^*\to c\bar{c}$.
For inclusive charm production, the virtual-photon contribution may have
a larger fraction than that in 
the inclusive light-hadronic decay
because the rate for $\Upsilon\to c\bar{c}gg$ is 
suppressed by order $\alpha_s$ compared to that for $\Upsilon\to ggg$.
We consider the virtual-photon contribution as well as the QCD
contributions from $b\bar{b}_1({}^3S_1)\to c\bar{c}gg$ and
$b\bar{b}_1({}^3S_1)\to c\bar{c}g\gamma$ modes.
In the current CLEO III analysis on the 
charmed-hadron ($h$) momentum distribution in $\Upsilon$ decay,
the virtual-photon contribution is subtracted experimentally~\cite{Roy}.
We therefore also present the results for the QCD contributions after 
excluding the virtual-photon process.

This paper is organized as follows.
In Sec.~\ref{sec:charm}, we present the NRQCD factorization
formulas for the inclusive charm production rate and charm 
momentum distribution in the spin-triplet $S$-wave bottomonium decay.
We also discuss the NRQCD matrix element that appear as a long-distance
factor in the factorization formula.
We calculate the charm-quark momentum distribution and the total
rate for inclusive charm production in the decay in Sec.~\ref{sec:rate}.
In Sec.~\ref{sec:meson},
in order to provide a theoretical prediction that can be compared
with CLEO III data, 
we illustrate the charmed-hadron momentum distributions 
which are obtained by convolving
the charm-quark momentum distribution
with fragmentation functions for $c\to h$ that have been fit to
$e^+e^-$ annihilation data. 
Finally, a brief summary of this work is given in Sec.~\ref{summary}.

\section{Charm quark production in $\bm{\Upsilon}$ 
decay\label{sec:charm}}

In this section, we summarize the NRQCD factorization formula for 
inclusive charm production in $\Upsilon$ decay. In many
aspects there is a large overlap with the formalism for calculating
the inclusive charm production rate in the $P$-wave bottomonium 
decay~\cite{Bodwin:2007zf}. In this work, we follow the same 
strategies that were employed in Ref.~\cite{Bodwin:2007zf}.
For details of the formalism, we refer the reader to 
Refs.~\cite{Bodwin:1994jh,Bodwin:2007zf}.

\subsection{NRQCD factorization formula\label{sec:NRQCD}}

At leading order in $v$, the NRQCD factorization formula 
for the inclusive charm production rate in $\Upsilon$ decay has 
an analogous form to that for the light-hadronic decay in 
Eq.~(\ref{Gam-ups}):
\begin{equation}
\Gamma[\Upsilon \to c + X] =
C_1^{(c)} \, 
\frac{\langle \mathcal{O}_1({}^3S_1) \rangle_\Upsilon}{m_b^2},
\label{Gam-c}%
\end{equation}
where $C_1^{(c)}$ is a dimensionless short-distance coefficient that 
depends on the mass ratio $m_c/m_b$ of the charm quark and the bottom
quark and $c + X$ represents all possible states containing a charmed
hadron, such as $D^+$, $D^0$, $D_s^+$, $\Lambda_c^+$, or their 
excited states.

\begin{figure}[tb]
\includegraphics*[width=8cm,angle=0,clip=true]{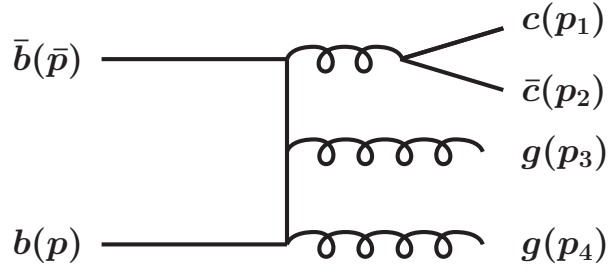}
\caption{
One of six Feynman diagrams for $b\bar{b} \to c\bar{c} g g$.
The other five diagrams are obtained by permuting the three gluons 
attached to the bottom-quark line.
}
\label{figFeyn}
\end{figure}

At leading order in $v$ and $\alpha_s$, the dominant source of 
$C_1^{(c)}$ is the decay of a $b\bar{b}_1(^3S_1)$ pair into $ggg^*$,
followed by $g^\ast \to c\bar{c}$. One of six Feynman diagrams
of the process is shown in Fig.~\ref{figFeyn} and the remaining five
diagrams are obtained by permuting the three gluons attached to the 
bottom-quark line.
The decay of a $b\bar{b}_1({}^3S_1)$ pair into $\gamma g g^\ast$,
followed by $g^\ast \to c\bar{c}$ also contributes to $C_1^{(c)}$.
We call the two processes involving $g^*\to c\bar{c}$ as QCD processes.

Another similar subprocesses that has the same final state $c\bar{c}gg$
as the QCD process is the decay of $b\bar{b}_1({}^3S_1) \to gg\gamma^\ast$
followed by $\gamma^\ast \to c\bar{c}$.
Because the $c\bar{c}$ pair created in this subprocess is in a color-singlet
state, there is no interference between this and the QCD process whose
$c\bar{c}$ pair is created in a color-octet state. The contribution of 
the subprocess $b\bar{b}_1({}^3S_1) \to gg\gamma^\ast$ followed by
$\gamma^\ast \to c\bar{c}$ to the $\Upsilon$ decay width is suppressed
to the QCD contribution by a multiplicative factor 
$8e_b^2 e_c^2 \alpha^2 N_c^2/\alpha_s^2/(N_c^2-4)$.
For $\alpha_s (m_\Upsilon/2)\approx 0.215$ 
and $\alpha(m_\Upsilon/2)\approx 1/132$,
the factor is about $ 0.09\%$.
Therefore, we neglect this subprocess.

In addition to the QCD modes, a virtual-photon mode, 
$b\bar{b}_1({}^3S_1)\to\gamma^*$,
followed by $\gamma^*\to c\bar{c}+X$ contributes to $C_1^{(c)}$.
A rough estimate of the branching fraction of the virtual-photon
channel is about 3\%, which is
the product of the measured leptonic width 
$\Gamma[\Upsilon \to e^+e^-]$, coupling $e_c^2$, and the color factor
$N_c$. This is comparable to the branching fraction 
$\textrm{Br}[\Upsilon\to c\bar{c}gg]$  predicted by Fritzsch and 
Streng~\cite{Fritzsch:1978ey}. 
In this work, we consider the virtual-photon contribution 
$C_1^{(c/\gamma^*)}$ from
$b\bar{b}_1({}^3S_1)\to \gamma^\ast \to c\bar{c}$
as well as the QCD contribution $C_1^{(c/g^*)}$
which is composed of 
$b\bar{b}_1({}^3S_1)\to c\bar{c}gg$ and
$b\bar{b}_1({}^3S_1)\to c\bar{c}g\gamma$.
Then the short-distance coefficient $C_1^{(c)}$ is expressed as
\begin{equation}
C_1^{(c)} = C_1^{(c/g^*)} + C_1^{(c/\gamma^*)}.
\label{c1sum}%
\end{equation}

The two QCD contributions to $C_1^{(c/g^*)}$ are essentially
the same except for overall factors so that
\begin{equation}
C_1^{(c/g^*)} = C_1^{(c\bar{c}gg)}+C_1^{(c\bar{c}g\gamma)}
= F_\gamma C_1^{(c\bar{c}gg)},
\label{c1c}%
\end{equation}
where 
$C_1^{(c\bar{c}gg)}$ and 
$C_1^{(c\bar{c}g\gamma)}$ are contributions of
$c\bar{c}gg$ and $c\bar{c}g\gamma$ channels to $C_1^{(c/g^*)}$, respectively.
The factor $F_\gamma$ is defined by 
\begin{equation}
F_\gamma = 1+ \frac{2e_b^2 \alpha}{\alpha_s} \frac{4 N_c}{(N_c^2-4)}. 
\label{F1}%
\end{equation}
For $\alpha_s(m_\Upsilon/2)\approx 0.215$ 
and $\alpha(m_\Upsilon/2)\approx 1/132$,
the numerical value for $F_\gamma$ is
about $1.02$, which indicates that the $c\bar{c}g\gamma$ contribution
is only about $2\%$ of the QCD contributions.
Since we are interested in the total decay  rate of $\Upsilon\to c+X$ 
and distributions of the charm quark with respect to its kinematic variables,
we integrate out the phase spaces for the gluons or the photon. Therefore,
the total and the differential width of the QCD mode
are the same as those for the $c\bar{c}gg$ final state
up to the multiplicative factor $F_\gamma$ in Eq.~(\ref{F1}).

As in the case of light hadronic decay of the $\Upsilon$,
inclusive charm production in $\Upsilon$ decay
may have contributions from the decay of the color-octet pair $b\bar{b}_8$
through $b\bar{b}_8 \to g^\ast$, followed by $g^\ast \to c\bar{c}$. 
While the color-octet contribution is suppressed to the color-singlet 
contribution by order $v^4$, the short-distance coefficient of the octet
process $b\bar{b}_8({}^3S_1)\to g^\ast\to c\bar{c}$ is enhanced by
$1/\alpha_s^2$. Therefore, the color-octet channel may have non-negligible
contributions, especially for the decay of higher resonances. 
In this work, we do not consider the color-octet contributions.

In a recent analysis being carried out
by the CLEO Collaboration, as a part of experimental measurement,
the virtual-photon contribution is subtracted from the data samples
of $\Upsilon\to c+X$ by scaling the continuum data by an extra factor
based on the branching fractions $\textrm{Br}[\Upsilon(nS)\to\mu^+\mu^-]$
and $R_{\textrm{hadrons}}$ but the data include the contribution
from $b\bar{b}_1(^3S_1) \to \gamma gg^*$ followed by
$g^* \to c \bar{c}$~\cite{Roy}.
The CLEO III data should, therefore, be
directly compared with the QCD contributions $C_1^{(c/g^*)}$ which
excludes  $C_1^{(c/\gamma^*)}$ from  $C_1^{(c)}$ in Eq.~(\ref{c1sum}).

In the remainder of this paper, we use the same conventions of 
superscripts to other variables as those used in the short-distance 
coefficients $C_1^{(c)}$, $C_1^{(c/g^*)}$, 
$C_1^{(c\bar{c}gg)}$, $C_1^{(c\bar{c}g\gamma)}$,
and $C_1^{(c/\gamma^*)}$. For example, 
$\Gamma^{(c)}=\Gamma[\Upsilon\to c+X]$ and
$\Gamma^{(c\bar{c}gg)}=\Gamma[\Upsilon\to c\bar{c}gg]$.

\subsection{Amplitude for $\bm{b}\bar{\bm{b}}$ annihilation into charm}

The short-distance coefficients $C_1^{(c)}$, $C_1^{(c/g^*)}$,
$C_1^{(c\bar{c}gg)}$, $C_1^{(c\bar{c}g\gamma)}$, and $C_1^{(c/\gamma^*)}$
are calculable by using perturbative matching, which involves
the computation of the amplitudes for the corresponding perturbative
short-distance processes such as $bb_1({}^3S_1)\to c\bar{c}gg$.
In this section, we summarize a way to calculate the
annihilation amplitude for the process $bb_1({}^3S_1)\to c\bar{c}gg$.
Computations of the amplitudes for the remaining processes are 
analogous to that for $bb_1({}^3S_1)\to c\bar{c}gg$.

At leading order in $\alpha_s$, the short-distance process 
for $\Upsilon\to c + X$ is
$b(p)\bar{b}(\bar{p}) \to c(p_1)\bar{c}(p_2) g(p_3) g(p_4)$
as shown in Fig.~\ref{figFeyn}. 
The momenta of the $b$ and the $\bar{b}$ can be expressed in terms of
the total momentum $P$ and the relative momentum
$q$ of the $b\bar{b}$ pair: 
\begin{subequations}
\begin{eqnarray}
p&=&\tfrac{1}{2}P+q,
\\
\bar{p}&=&\tfrac{1}{2}P-q,
\end{eqnarray}
\end{subequations}
where the $p$ and the $\bar{p}$ satisfy the on-shell conditions
$p^2=\bar{p}^2=m_b^2$ and $P\cdot q=0$.
In the rest frame of the $b\bar{b}$ pair, 
$P=(2E_b,0)$ and $q=(0,\bm{q})$, where
$E_b=\sqrt{m_b^2+\bm{q}^2}$. 
In general, the perturbative amplitude for the $b\bar{b}$ annihilation
process is expressed as
\begin{equation}
\bar{v}(\bar{p}) \mathcal{M}[b\bar{b}]  u(p) =
\textrm{Tr} \big(\mathcal{M}[b\bar{b}]  u(p) \bar{v}(\bar{p}) \big),
\label{vbarMu}%
\end{equation}
where $\mathcal{M}[b\bar{b}]$ is a matrix that acts on spinors with 
both Dirac and color indices. The matrix $\mathcal{M}[b\bar{b}]$ 
for the short-distance process 
$b\bar{b}\to c(p_1)\bar{c}(p_2) g(p_3) g(p_4)$ is given by
\begin{eqnarray}
\mathcal{M}[b \bar{b}] &=&
\frac{16 \pi^2 \alpha_s^2}{(p_1+p_2)^2}
  \bar{u}(p_1) T^a \gamma_\lambda v(p_2)
  \epsilon^{b *}_{1\,\sigma}(p_3)\epsilon^{c *}_{2\,\tau}(p_4)
\nonumber
\\
&& 
\times 
\sum_\textrm{perm}
\left[ \gamma^\lambda \Lambda(p-p_3-p_4) \gamma^\sigma
\Lambda(p-p_4) \gamma^\tau 
\otimes T^a T^b T^c \right],
\label{amp-bb}%
\end{eqnarray}
where $\sum_\textrm{perm}$ means the summation over the permutations
of the three gluons attached to the bottom-quark line.
$T^a$ is a generator of the fundamental representation 
for the SU(3) color group and $a$,$b$, and $c$ are color indices
for the gluons. $\epsilon_1$ and $\epsilon_2$ are polarization vectors
for the external gluons with momenta $p_3$ and $p_4$, respectively.
Note that the expression in Eq.~(\ref{amp-bb}) is valid to any order
in $v$. The function $\Lambda(k)$ is defined by
\begin{equation}
\Lambda(k)=\frac{/\!\!\!k+m_b}{k^2-m_b^2}.
\end{equation}

In general, the amplitude (\ref{vbarMu}) contains contributions other
than the color-singlet $b\bar{b}_1({}^3S_1)$ state, which we want to 
project out. A convenient way to carry out the projection is
to replace the spinor product $u(p)\bar{v}(\bar{p})$ in 
Eq.~(\ref{vbarMu}) by the direct product of the color-singlet projection
operator $\pi_1$ and the spin-triplet projector 
$\epsilon_\mu\Pi^\mu_3$~\cite{Kuhn:1979bb,Guberina:1980dc,Bodwin:2002hg},
where
\begin{subequations}
\begin{eqnarray}
\pi_1 &=& \frac{1}{\sqrt{N_c}} \mathbbm{1},
\label{color-projector}%
\\
\Pi_3^\mu &=&
-\frac{1}{4\sqrt{2}E_b(E_b+m_b)}
(/\!\!\!{p}+m_b)(\,/\!\!\!\!P\!+\!2E_b) \gamma^\mu
(/\!\!\!\bar{p}-m_b) ,
\label{spin-projector}%
\end{eqnarray}
\label{projectors}%
\end{subequations}
where $\mathbbm{1}$ is 
the SU(3) color unit matrix and
$\epsilon$ is the polarization four-vector of the $b\bar{b}_1({}^3S_1)$
state so that $P \cdot \epsilon  = 0$. The projectors (\ref{projectors})
are normalized as $\textrm{Tr}[\pi_1\pi_1^\dagger]=1$ and
$\textrm{Tr}[(\epsilon\cdot\Pi_3)%
(\epsilon\cdot\Pi_3)^\dagger]=4p_0\bar{p}_0$. At leading order in $v$, 
the amplitude for the color-singlet spin-triplet $S$-wave 
$b\bar{b}$ pair can be written as
$\epsilon_\mu \mathcal{A}_1^\mu[b\bar{b}_1(^3S_1)]$, where
\begin{equation}
\mathcal{A}^\mu_1 =
\textrm{Tr} \big( \mathcal{M}[b\bar{b}]\,
\Pi_3^\mu \otimes \pi_1  \big) 
\Big|_{q=0}.
\label{amp-S}%
\end{equation}
Because we are working in the leading order in $v$,
we put $q=0$ and, therefore, $E_b=m_b$. 

The amplitude (\ref{amp-S}) is finite in the soft
limits of any external gluons. At higher orders in $v$, infrared divergences 
arise in this $S$-wave amplitude while a $P$-wave amplitude has an infrared
divergence at leading order in $v$.
The amplitude (\ref{amp-S}) is sensitive to 
the ratio $m_c/m_b$. In the massless charm-quark limit $m_c/m_b\to 0$,
the amplitude (\ref{amp-S}) acquires a collinear divergence, which cancels
that arising from the charm-quark loop corrections to the gluon wave function
for the $\Upsilon\to ggg$ process~\cite{Bodwin:2007zf}.
Because the amplitude (\ref{amp-S}) is free of any infrared and
collinear divergences for $m_c \neq 0$ and $q=0$, 
we do not need any regularization scheme and
work in four space-time dimensions.

\subsection{Short-distance coefficients}

We proceed to calculate $C^{(c)}_1$  in the NRQCD factorization formula 
(\ref{Gam-c}). The short-distance coefficient $C_1^{(c)}$ is
insensitive to the long-distance nature of the $\Upsilon$. 
Therefore, $C_1^{(c)}$ is calculable perturbatively. This 
can be done by perturbative matching~\cite{Bodwin:1994jh,Bodwin:2007zf}.
In order to determine the short-distance coefficients $C_1^{(c)}$,
we must calculate the annihilation rate for the color-singlet
spin-triplet $b\bar{b}$ state by using perturbative QCD.
The perturbative analog of the NRQCD factorization formula in
Eq.~(\ref{Gam-c}) for the annihilation rate of the
$b\bar{b}_1({}^3S_1)$ pair is
\begin{equation}
d\Gamma[b\bar{b}_1({}^3S_1)\to c+X]=dC_1^{(c)}
\frac{\langle \mathcal{O}_1({}^3S_1)\rangle_{b\bar{b}_1({}^3S_1)}
     }
{m_b^2},
\label{dGam-c-bb}%
\end{equation}
where the factorization formula (\ref{dGam-c-bb}) is written in
a differential form. This form is useful for our purpose of
calculating the momentum distribution of the charm quark.

\subsubsection{Calculation of $C_1^{(c/g^*)}$}
The differential annihilation rate of a color-singlet spin-triplet $S$-wave
$b\bar{b}$ state into charm through the process 
$b\bar{b}_1({}^3S_1)\to c\bar{c}gg$ can be expressed as
\begin{equation}
d\Gamma[b\bar{b}_1({}^3S_1)\to c\bar{c}gg]=
\left(
\frac{1}{3}I_{\mu\nu}
\sum_{c\bar{c}gg}\mathcal{A}_1^{\mu}\mathcal{A}_1^{\nu*}
\right)\frac{d\Phi_4}{2!},
\label{dGam-c-bb-pert}%
\end{equation}
where 
$\mathcal{A}_1$ is the perturbative amplitude (\ref{amp-S})
for $b\bar{b}_1({}^3S_1)\to c\bar{c}gg$,
$I^{\mu\nu}$ is the spin-$1$ polarization tensor 
for the $b\bar{b}_1({}^3S_1)$,
\begin{equation}
I^{\mu\nu}=-g^{\mu\nu}+\frac{P^\mu P^\nu}{P^2},
\label{Imunu}%
\end{equation}
$d\Phi_4$ is the four-body phase space for $c\bar{c}gg$,
and $\sum_{c\bar{c}gg}$ indicates summation over the spin states 
of $c\bar{c}gg$.
The factor $1/3$ in Eq.~(\ref{dGam-c-bb-pert}) comes from
averaging over the spin states for the $b\bar{b}_1({}^3S_1)$.
A factor of $1/2!$ is multiplied to the four-body phase space
because there are two identical particles in the $c\bar{c}gg$ 
final state, whose phase spaces are integrated out.

In order to complete the matching calculation for 
$dC^{(c)}_1$, we need to compute the perturbative NRQCD 
matrix element:
\begin{equation}
\langle\mathcal{O}_1({}^3S_1) 
\rangle_{b\bar{b}_1({}^3S_1)}
=2N_c (2E_b)^2= 8N_cm_b^2+\mathcal{O}(v^2).
\label{me-pert-norm}%
\end{equation}
Substituting Eqs.~(\ref{dGam-c-bb-pert}) and (\ref{me-pert-norm})
into Eq.~(\ref{dGam-c-bb})
and multiplying by $F_\gamma$ in order to take into account 
the $c\bar{c}g\gamma$ process as well as $c\bar{c}gg$,
we find the differential 
short-distance coefficient $dC_1^{(c/g^*)}$:
\begin{equation}
dC_1^{(c/g^*)}=
\frac{F_\gamma}
     {8N_c}
\left(
\frac{1}{3}I_{\mu\nu}
\sum_{c\bar{c}gg}\mathcal{A}_1^{\mu}\mathcal{A}_1^{\nu*}
\right)\frac{d\Phi_4}{2!}.
\label{dC}%
\end{equation}
If we replace $F_\gamma$ in Eq.~(\ref{dC}) with unity or
$F_\gamma-1$, we get 
the expression for $dC_1^{(c\bar{c}gg)}$ or $dC_1^{(c\bar{c}g\gamma)}$,
respectively.

A parameterization of the four-body phase space $d\Phi_4$
for the $c\bar{c}gg$ final state is derived in Appendix \ref{app}:
\begin{equation}
d\Phi_4
=
\frac{E_b^4}{2^{12} \pi^7}
\frac{r_Y (x_1^2-r_c)^{1/2} \lambda^{1/2}(r_X^2, r_Y^2, r_c)}{r_X^2}
dx_1 dr_Y d\Omega_2^* d\Omega_3^*,
\label{dPhi4}%
\end{equation}
where $\lambda(a,b,c)=a^2+b^2+c^2-2 a b - 2 b c - 2 c a$.
In Eq.~(\ref{dPhi4}),  we have not set $v=0$  so that
the expression can be used for a more general case.
$x_1$, $r_Y$, and $r_X$ in Eq.~(\ref{dPhi4})
are dimensionless variables defined by
\begin{subequations}
\begin{eqnarray}
x_1 &=& E_1/E_b, \\
r_Y &=& m_Y/E_b, \\
r_X &=& m_X/E_b, 
\end{eqnarray}
\label{xi-definition}%
\end{subequations}
where $E_1$ is the energy of the charm quark in the rest frame of 
the $b\bar{b}$ pair and the invariant masses $m_X$ and $m_Y$ are defined 
by $m_X^2=X^2=(p_2+p_3+p_4)^2$ and $m_Y^2=Y^2=(p_3+p_4)^2$, respectively.
The ranges of the integration variables are
\begin{subequations}
\begin{eqnarray}
\label{x1-bound}%
\sqrt{r_c}\leq &x_1& \leq1
,
\\
\label{xY-bound}%
0 \leq &r_Y& \leq \sqrt{4 - 4x_1 + r_c}-\sqrt{r_c},
\end{eqnarray}
\label{bound}%
\end{subequations}
where 
\begin{equation}
r_c=m_c^2/E_b^2.
\label{rc}%
\end{equation}
$d\Omega_2^*$ and $d\Omega_3^*$ are solid-angle elements of
the charm antiquark with momentum $p_2$ in the $X$-rest frame and
the gluon with momentum $p_3$ in the $Y$-rest frame,
respectively. 

\subsubsection{Calculation of $C_1^{(c/\gamma^*)}$}
The virtual-photon ($\gamma^*$) contribution  
$C^{(c/\gamma^*)}_1$ to the short-distance coefficient $C^{(c)}_1$ 
in Eq.~(\ref{c1c})
is proportional to that for the leptonic decay of the $\Upsilon(nS)$:
\begin{equation}
dC_1^{(\ell^+\ell^-)}=
\frac{2\pi}{3}e_b^2\alpha^2
\delta(1-x_1)dx_1.
\end{equation}
The short-distance coefficient $C^{(c/\gamma^*)}_1$ is
\begin{eqnarray}
dC_1^{(c/\gamma^*)}&=&e_c^2 N_c \left( 1+\frac{r_c}{2}\right)
                  \sqrt{1-r_c}\times dC_1^{\ell^+\ell^-}
\nonumber\\
&=&
\frac{\pi}{3}e_b^2e_c^2N_c\alpha^2 (2+r_c)
\sqrt{1-r_c}\,\delta(1-x_1)dx_1,
\label{dcgamma}%
\end{eqnarray}
where the factor $\sqrt{1-r_c}$ is the ratio of the 
phase space for the $c\bar{c}$ final state to 
the massless two-body phase space.
The QED coupling $\alpha(\mu)$ in Eq.~(\ref{dcgamma}) is defined by
the running coupling constant $\alpha(m_\Upsilon)=1/131$, $1/130.9$, and
$1/130.9$ for the radial quantum number $n=1$, $2$, and $3$, respectively.

\subsection{NRQCD matrix elements
\label{sec:matrix}}
In order to predict the decay widths (\ref{Gam-c}) for the radial
quantum numbers $n=1$, 2, and 3,
we must know the numerical value of 
$\langle \mathcal{O}_1 \rangle_{\Upsilon}$ for each state.
In principle, the NRQCD matrix elements may be calculated 
in lattice simulations~\cite{Bodwin:1996tg,Bodwin:2001mk}.
However, the only available NRQCD matrix element for 
an $S$-wave bottomonium
is that for the $1S$ state~\cite{Bodwin:2001mk}.
Extrapolating the value in Ref.~\cite{Bodwin:2001mk},
which is obtained from an unquenched calculation 
using two dynamical light quarks, to three light-quark flavors,
one obtains $\langle \mathcal{O}_1%
\rangle_{\Upsilon(1S)} = 3.95 \pm 0.43~\textrm{GeV}^3$,
where we use the same extrapolation method that is given in 
Ref.~\cite{Bodwin:2007zf}.

The NRQCD matrix element $\langle \mathcal{O}_1 \rangle_\Upsilon$
may also be determined by comparing the NRQCD factorization formula
for $\Gamma[\Upsilon\to e^+ e^-]$ with the empirical values,
which are measured with uncertainties of order $2\%$.
Recently, a method to resum a class of relativistic corrections to 
$S$-wave quarkonium processes
has been developed~\cite{Bodwin:2006dn,Bodwin:2007fz},
where order-$\alpha_s$ and resummed relativistic corrections are included
in the NRQCD factorization formula for the leptonic decay width.
This method has been used to determine the color-singlet NRQCD 
matrix elements for the $S$-wave charmonium 
states~\cite{Chung:2007ke,Bodwin:2007fz}
and to the calculation of exclusive two-charmonium 
production in $e^+e^-$ annihilation~\cite{Bodwin:2006ke,Bodwin:2007ga}.
Using the method given in Ref.~\cite{Bodwin:2007fz}, we can determine
the NRQCD matrix elements for each $nS$ state:
\begin{subequations}
\begin{eqnarray}
\langle \mathcal{O}_1 \rangle_{\Upsilon(1S)} &=&
3.07^{+0.21}_{-0.19}  ~ \textrm{GeV}^3, 
\label{ME1S}%
\\
\langle \mathcal{O}_1 \rangle_{\Upsilon(2S)} &=&
1.62^{+0.11}_{-0.10} ~ \textrm{GeV}^3, \\
\langle \mathcal{O}_1 \rangle_{\Upsilon(3S)} &=&
1.28^{+0.09}_{-0.08} ~ \textrm{GeV}^3.
\end{eqnarray}
\label{ME}%
\end{subequations}
The uncertainties in the matrix elements in Eq.~(\ref{ME}) reflect 
the errors in the lattice string tension, the one-loop pole mass of the 
bottom quark, the strong coupling constant, and the measured values for
the leptonic widths as well as corrections of order $v^2$ that are not
included in the potential model~\cite{Bodwin:2007fz}.
The central value of the matrix element (\ref{ME1S}) for $\Upsilon(1S)$ 
is smaller than the lattice estimate by about 22\%.
In our numerical calculations, we use the values for the NRQCD matrix
elements given in Eq.~(\ref{ME}).

\section{Decay rate}
\label{sec:rate}

\begin{figure}[tb]
\includegraphics*[width=12cm,angle=0,clip=true]{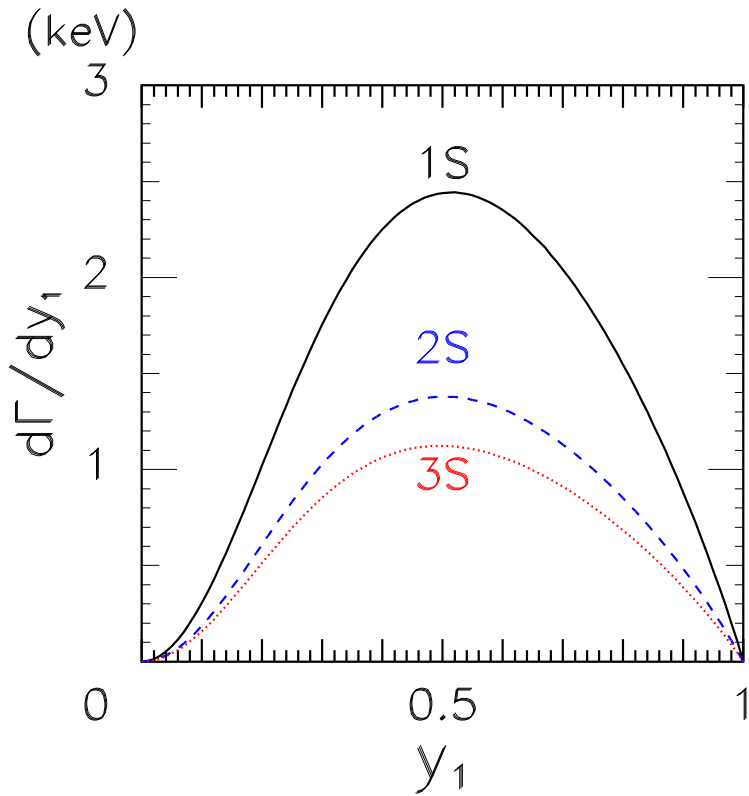}
\caption{
Distributions of the scaled momentum fraction $y_1$ for the charm quark 
in decays of $\Upsilon(nS)$ for $n=1$ (solid line), 2 (dashed line), 
and 3 (dotted line) by using $m_b=4.6$ GeV, 
$\alpha_s(m_\Upsilon/2)=0.215$, $0.212$, and $0.210$, and
$\langle \mathcal{O}_1 \rangle_{\Upsilon} = 3.07$, $1.62$, and $1.28$ 
$\textrm{GeV}^3$ for the $1S$, $2S$, and $3S$ states, respectively.
The distributions in the range $y_1<1$ are for the QCD contribution 
$d\Gamma^{(c/g^*)}$. There is a delta function at $y_1=1$ from the
virtual-photon contribution 
$d\Gamma^{(c/\gamma^*)}$ in Eq.~(\ref{dcgamma}),
which is not shown in this figure.
}
\label{fig2}%
\end{figure}

\subsection{Charm-quark momentum distribution}

In the quarkonium rest frame,
the NRQCD factorization formula for the differential distribution
with respect to the charm-quark energy fraction $x_1$ in 
$\Upsilon$ decay is easily deduced from Eq.~(\ref{Gam-c}):
\begin{equation}
\frac{d\Gamma^{(c)}}{dx_1} =
\frac{dC_1^{(c)}}{dx_1} \,
\frac{\langle \mathcal{O}_1 \rangle_{\Upsilon}}{m_b^2},
\label{dGamdx1}%
\end{equation}
where the differential coefficient is defined by 
$dC_1^{(c)}=dC_1^{(c/g^*)}+dC_1^{(c/\gamma^*)}$, which is analogous
to Eq.~(\ref{c1sum}), and $dC_1^{(c/g^*)}$ and $dC_1^{(c/\gamma^*)}$
are given in Eqs.~(\ref{dC}) and (\ref{dcgamma}).
Integration over the variables $dr_Y$, $d\Omega_2^*$,
and $d\Omega_3^*$ is implicit in $dC_1^{(c/g^*)}\!/dx_1$ 
on the right side of Eq.~(\ref{dGamdx1}).

In the $\Upsilon$ rest frame, the magnitude of the three-momentum
of the charm quark is determined by the energy fraction $x_1$ as
Eq.~(\ref{absp1}). Using this relation, we obtain the 
charm-quark momentum distribution from Eq.~(\ref{dGamdx1}).
It is convenient to introduce the scaled momentum fraction $y_1$ 
that is defined as the momentum of the charm quark divided by
the maximum value that is kinematically allowed
in  $\Upsilon$ decay~\cite{Bodwin:2007zf}. The physical range
of the variable is simple: $0 < y_1 < 1$.
By using the relations between $x_1$ and $y_1$
\begin{subequations}
\begin{eqnarray}
x_1&=&\sqrt{(1-r_c) y_1^2+r_c},
\label{x1inverse}%
\\
y_1 &=& \sqrt{\frac{x_1^2 - r_c}{1-r_c}},
\label{y1-def}%
\end{eqnarray}
\end{subequations}
we obtain the distribution for the scaled momentum fraction $y_1$:
\begin{equation}
\frac{d\Gamma^{(c)}}{dy_1} =
\frac{(1-r_c) y_1}{\sqrt{(1-r_c)y_1^2 + r_c}} \, 
\frac{d\Gamma^{(c)}}{dx_1},
\label{dGy1}%
\end{equation}
where $d\Gamma^{(c)}/dx_1$ is given in Eq.~(\ref{dGamdx1}).

In the NRQCD factorization formula (\ref{dGamdx1}), 
the bottom-quark mass $m_b$ appears in common for all three
$S$-wave states. In order to be consistent with a preceding 
work~\cite{Bodwin:2007zf}, we use the one-loop pole mass
$m_b=4.6$ GeV for that $m_b$. 
The dimensionless short-distance coefficient 
$dC_1^{(c)}/dx_1$ depends on the strong coupling $\alpha_s$ and the ratio $r_c$.
For the strong coupling, we use 
$\alpha_s(m_\Upsilon/2) = 0.215$, $0.212$, and $0.210$ for the ratial 
quantum number $1$, $2$, and $3$ of the $\Upsilon$, respectively.
The ratio $r_c$ depends on $m_c/E_b$. In the nonrelativistic limit $v\to 0$,
which we are taking in this work, $r_c\to r\equiv (m_c/m_b)^2$.
Then the bounds (\ref{bound}) of the variables $x_1$ and $r_Y$ are determined 
by $r$. 
We take $r_c=r=4 m_D^2/m_{\Upsilon}^2$, where $m_D=1.87$ GeV is 
the average of the masses of the $D^0$ and $D^+$. 
For the $S$-wave spin-triplet bottomonium masses, we use
$m_{\Upsilon}=9.46$ GeV, $10.02$ GeV,
and $10.36$ GeV, for the radial quantum number $n=1$, 
2, and 3, respectively~\cite{Yao:2006px}.
This choice of $r_c$ correctly reflects the physical 
kinematics~\cite{Bodwin:2007zf}.
We take the numerical values in Eq.~(\ref{ME}) for the color-singlet 
NRQCD matrix elements.

Our theoretical prediction for the momentum distribution of the
charm quark in the inclusive $\Upsilon$ decay is shown in 
Fig.~\ref{fig2} in terms of $d\Gamma^{(c/g^*)}/dy_1$, where
the solid, dashed, and dotted lines are
the distributions for the $1S$, $2S$, and $3S$ states, respectively.
Because the color-singlet $S$-wave amplitudes for the processes
are infrared finite, the distributions in Fig.~\ref{fig2}
are finite over the whole range of $y_1$. This is different from
those for $\chi_{bJ}\to c+X$  $(J=0,1,2)$, which grow rapidly
as $y_1\to 1$~\cite{Bodwin:2007zf}.
The $y_1$ distributions are broad because of the four-body nature of the
$c\bar{c}gg$ ($c\bar{c}g\gamma$)  final state. 
The curves have the maximum values
$2.44$ keV at $y_1=0.51$,
$1.38$ keV at $y_1=0.50$, and
$1.12$ keV at $y_1=0.50$ for 
$\Upsilon{(1S)}$,
$\Upsilon{(2S)}$, and
$\Upsilon{(3S)}$, respectively.

As we have stated in Sec.~\ref{sec:NRQCD}, the distributions in
Fig.~\ref{fig2} include only 
$b\bar{b}_1({}^3S_1)\to c\bar{c} gg$ and
$b\bar{b}_1({}^3S_1)\to c\bar{c} g \gamma $ contributions to
the short-distance coefficient $dC^{(c)}_1$ (\ref{dC}).
For $y_1<1$, the leading virtual-photon contribution 
$\Upsilon\to\gamma^*\to c\bar{c}$ does not contribute because
the distribution is proportional to $\delta(1-y_1)$.
The sharp peak from the virtual-photon contribution 
$d\Gamma^{(c/\gamma^*)}\!/dy_1$ at the end
point has contributions to the total inclusive charm production
rate comparable to that of
the QCD contributions illustrated in Fig.~\ref{fig2}. 

\begin{table}[t]
\caption{\label{table:decayrate}%
Inclusive charm production rate $\Gamma^{(c)}$ and
partial widths $\Gamma^{(c/g^*)}$ and $\Gamma^{(c/\gamma^*)}$
in units of keV for 
$m_b=4.6 \pm 0.1$ GeV, and
$\langle \mathcal{O}_1 \rangle_{\Upsilon}$ in Eq.~(\ref{ME}).
Uncertainties are estimated as stated in the text.
The partial widths $\Gamma^{(c\bar{c}gg)}$ and
$\Gamma^{(c\bar{c}g\gamma)}$ can be obtained by multiplying
$\Gamma^{(c/g^*)}$ by factors $F^{-1}_\gamma\approx 0.982$ and 
$1-F^{-1}_\gamma\approx 0.0184$, respectively.
}
\begin{ruledtabular}
\begin{tabular}{lccc}
state $\backslash$ $\Gamma$ (keV)
 & $\Gamma^{(c/g^*)}$
 & $\Gamma^{(c/\gamma^*)}$
 & $\Gamma^{(c)}$
\\
\hline
$\Upsilon(1S)$
& $1.47\pm 0.36$ 
& $2.60\pm 0.65$ 
& $4.07\pm 0.75$ 
\\
$\Upsilon(2S)$
& $0.83\pm 0.20$
& $1.38\pm 0.34$
& $2.21\pm 0.40$
\\
$\Upsilon(3S)$
& $0.68\pm 0.16$
& $1.09\pm 0.27$
& $1.77\pm 0.32$
\end{tabular}
\end{ruledtabular}
\end{table}
\begin{table}[t]
\caption{\label{table:br}%
Branching fractions Br$^{(i)}=\Gamma^{(i)}/\Gamma[\Upsilon]$ for
$(i)=(c/g^*)$, $(c/\gamma^*)$, and $(c)$, where 
$\Gamma^{(i)}$'s are given in Table~\ref{table:decayrate} and
$\Gamma[\Upsilon]$ is the measured total width~\cite{Yao:2006px} 
of $\Upsilon$ for radial quantum numbers $n=$1, 2, and 3.
}
\begin{ruledtabular}
\begin{tabular}{lccc}
state $\backslash$ Br (\%)
 & Br$^{(c/g^*)}$
 & Br$^{(c/\gamma^*)}$
 & Br$^{(c)}$
\\
\hline
$\Upsilon(1S)$
& $2.72\pm 0.67$
& $4.81\pm 1.21$
& $7.53\pm 1.39$
\\
$\Upsilon(2S)$
& $2.60\pm 0.67$
& $4.30\pm 1.12$
& $6.90\pm 1.37$
\\
$\Upsilon(3S)$
& $3.34\pm 0.87$
& $5.36\pm 1.41$
& $8.70\pm 1.74$
\end{tabular}
\end{ruledtabular}
\end{table}

\subsection{Total charm production rate}
\label{sec:totalrate}%

The inclusive charm production rate in $\Upsilon$ decay
can be calculated by integrating the differential
rate~(\ref{dGamdx1}) over $x_1$ or the differential rate (\ref{dGy1})
over $y_1$. In Table~\ref{table:decayrate} we list 
the partial widths $\Gamma^{(c/g^*)}$ and $\Gamma^{(c/\gamma^*)}$ and
the total width $\Gamma^{(c)}$. The partial width $\Gamma^{(c/g^*)}$
for the QCD process is the sum of $\Gamma^{(c\bar{c}gg)}$ and
$\Gamma^{(c\bar{c}g\gamma)}$, whose values can be obtained by
multiplying  $\Gamma^{(c/g^*)}$ by factors $F^{-1}_\gamma\approx 0.982$ and
$1-F^{-1}_\gamma\approx 0.0184$, respectively.
The theoretical uncertainties in Table~\ref{table:decayrate} are
estimated as follows.
We consider the uncertainties of the pole
mass appearing in the factorization formula (\ref{Gam-c})
as $m_b=4.6\pm 0.1$~GeV. As we have discussed earlier, the $m_c$ dependence
of the dimensionless short-distance coefficients are completely determined
by $r_c=r=4m_D^2/m_\Upsilon^2$. We do not vary $r$ because the measured
values for the hadron masses do not contribute to errors at the level of 
accuracy we take into account.
We use the uncertainties in Eq.~(\ref{ME}) for the NRQCD matrix elements.
We consider the errors from uncalculated order-$v^2$ and order-$\alpha_s$ 
corrections by multiplying the central values shown in 
Table~\ref{table:decayrate} by $v^2\approx 10$\% and 
by $\alpha_s(m_\Upsilon/2)\approx 0.215$,
respectively. The error bars in the widths appearing in
Table~\ref{table:decayrate} are obtained by combining the uncertainties
that are listed above in quadrature.

The branching fractions for the decay channels 
$c\bar{c}gg+c\bar{c}g\gamma$, $\gamma^*\to c\bar{c}$, and $c+X$
are listed in Table~\ref{table:br}. The numbers are obtained by
dividing the numbers in  Table~\ref{table:decayrate} by the
measured widths
$\Gamma[\Upsilon(1S)]= 54.02\pm 1.25$ keV,
$\Gamma[\Upsilon(2S)]= 31.98\pm 2.63$ keV, and
$\Gamma[\Upsilon(3S)]= 20.32\pm 1.85$ keV~\cite{Yao:2006px}.  From 
Tables~\ref{table:decayrate} and \ref{table:br}, we conclude
that the virtual-photon contribution is actually greater than
the QCD contribution for all three $S$-wave states.

At the next-to-leading order in $\alpha_s$, the virtual-photon
contribution may contribute to the region $y_1<1$ because of
real-gluon emissions. Because the leading-order
contribution to $\Gamma^{(c/\gamma^*)}$ is greater than
$\Gamma^{(c/g^*)}$ by about factors of $1.6\,$--$\,1.8$, the 
order-$\alpha_s$ corrections to $\Gamma^{(c/\gamma^*)}$ may modify 
the shape of the $y_1$ distributions for $y_1<1$ by about $30$\%.

\section{Charmed-hadron momentum distribution}
\label{sec:meson}%

In Sec.~\ref{sec:rate}, we calculated the momentum distribution
of the charm quark and the inclusive charm production rate
in $\Upsilon$ decay. 
Since the charm quark hadronizes into one of charmed hadrons
with a probability of almost 100\%, the charm production rate can be 
interpreted to be the sum of the production rates for the charmed hadrons $h$.
The charmed hadrons
include the $D^0$, $D^+$, $D^+_s$, and $\Lambda^+_c$, which are stable 
under strong and electromagnetic interactions, and excited
charmed hadrons, whose decay product includes $D^0$, $D^+$, $D^+_s$, or 
$\Lambda^+_c$. As is discussed in Ref.~\cite{Bodwin:2007zf},
the momentum distribution of a charmed hadron produced in $\Upsilon$
decay is softer than that of the charm, because of the effect of
hadronization. The momentum distribution for a charmed hadron $h$ can be
obtained by convolving the charm-momentum distribution with a fragmentation
function for the charm quark to fragment into a $h$. For more details,
we refer the reader to Ref.~\cite{Bodwin:2007zf} and references therein.

The fragmentation function $D_{c\to h}(z)$ gives the probability 
density for a charm quark with plus component of light-cone momentum
$E_1+p_1$ to hadronize into a charmed hadron $h$ with 
light-cone momentum $E_h+p_h=z(E_1+p_1)$. The fraction $z$ can be
expressed in terms of scaled light-cone momentum fractions
$z_1$ for the charm and $z_h$ for the charmed hadron, which are
analogous to the scaled momenta $y_1$ and 
$y_h$~\cite{Bodwin:2007zf}, where $z_1$ is
\begin{equation}
z_1 = \frac{\sqrt{(1-r_c)y_1^2 + r_c} + \sqrt{1-r_c} \, y_1}
           {1 + \sqrt{1-r_c}}.
\end{equation}
Then, the fraction $z$ is expressed as
\begin{equation}
z = \frac{z_h}{z_1}\times 
\frac{\left. (E_h+p_{h})\right|_\textrm{max}}
     {\left. (E_1+p_{1})\right|_\textrm{max}},
\label{z}%
\end{equation}
where the last factor on the right side of Eq.~(\ref{z}) becomes
unity if the difference between the mass of the charm quark
and that of the charmed hadron can be neglected.
Within this approximation the momentum distribution of 
the charmed hadron can be written as~\cite{Bodwin:2007zf}
\begin{eqnarray}
\frac{d\Gamma}{d y_h}
&=&
\frac{dz_h}{dy_h} \int_{z_h}^{1} \frac{dz_1}{z_1} \,
D_{c\to h}(z_h/z_1)\, \frac{dy_1}{dz_1}
\frac{d\Gamma}{d y_1}
\nonumber\\
&=&
\frac{\sqrt{1-r_c}}{\sqrt{(1-r_c)y_h^2+r_c }}
\int_{y_h}^1 dy_1
  \mathcal{D}_{c\to h}\left(
  \frac{\sqrt{(1-r_c)y_h^2+r_c} +\sqrt{1-r_c}y_h }
       {\sqrt{(1-r_c)y_1^2+r_c} +\sqrt{1-r_c}y_1 }
\right) \frac{d\Gamma}{dy_1},
\label{dGamD}%
\end{eqnarray}
where $\mathcal{D}_{c\to h}(z)= z D_{c\to h}(z)$.

Explicit parameterizations for $D_{c\to h}(z)$ for some
charmed hadrons $h$ 
which are currently available can be found, for example, in 
Refs.~\cite{Artuso:2004pj,Seuster:2005tr}.
Following a previous work on the charm production in the $P$-wave 
bottomonium decay~\cite{Bodwin:2007zf}, we refer to the results 
obtained by the Belle Collaboration~\cite{Seuster:2005tr}. They
determined the optimal values of the parameters for analytic
parameterizations of $D_{c\to h}(z)$ for various charmed hadrons
by comparing their measured momentum distribution in $e^+e^-$ 
annihilation near the center-of-momentum energy $10.6$ GeV with
the distributions predicted by Monte Carlo generators and fragmentation
functions~\cite{Seuster:2005tr}.

In our numerical analysis, we consider all the charmed hadrons considered
in Ref.~\cite{Seuster:2005tr}. These cover all the channels being
analyzed by the CLEO Collaboration: 
$\Upsilon(1S)\to h+X$, where $h=D^0$, $D^+$, $D^+_s$, $D^{*+}$, and 
$\Lambda_c^+$. We use the Kartvelishvili-Likhoded-Petrov (KLP) 
fragmentation function~\cite{Kartvelishvili:1977pi}, which was used
in the analysis of charmed-hadron momentum distribution in $\chi_b$
decays. The KLP fragmentation function has a simple parameterization
depending only on the light-cone momentum fraction $z$:
\begin{equation}
D_{c\to h}(z)= N_h z^{\alpha_c} (1-z).
\label{fragment}%
\end{equation}
The optimal values for the parameter $\alpha_c$ determined by the Belle 
Collaboration are $\alpha_c=4.6$, $4$, $5.6$, $5.6$, and $3.6$ for 
$D^0$, $D^+$, $D^+_s$, $D^{*+}$, and $\Lambda^+_c$, 
respectively~\cite{Seuster:2005tr}.

The normalization $N_h$ is determined by the constraint
$\int_0^1 dz D_{c\to h}(z)=$Br$[c\to h]$.
Using Table~X of Ref.~\cite{Seuster:2005tr},
one can infer that the inclusive fragmentation probabilities are
Br$[c\to h]=0.565$, $0.268$, $0.092$, $0.220$, and $0.075$
for the $D^0$, $D^+$, $D^+_s$, $D^{*+}$, and $\Lambda^+_c$, respectively.
As a result, we determine the normalization factors for various
charmed hadrons as $N_{D^0}=20.9$, $N_{D^+}=8.04$, $N_{D_s^+}=4.59$, 
$N_{D^{*+}}=11.0$, and $N_{\Lambda_c^+}=1.93$. 
Note that the branching fractions for the $c\to D^0$ and $c\to D^+$
include feeddowns from higher resonances $D^{0*}$ and
$D^{*+}$.
Substituting Eq.~(\ref{fragment}) into Eq.~(\ref{dGamD}) and using
the parameters listed above, we evaluate the momentum distributions
for the charmed hadrons. 
In Eq.~(\ref{dGamD}), we use $r_c=r=4 m_h^2/m_\Upsilon^2$ for $z_h$ where  
$h=D_s^+$, $D^{* +}$, and $\Lambda_c^+$
while $r_c=r=4 m_D^2/m_\Upsilon^2$ for $z_1$ and $z_D$
where $D=D^0$ and $D^+$.
We use $m_{D_s^+}=1.97$ GeV, $m_{D^{*+}}=2.01$ GeV, and 
$m_{\Lambda_c^+}=2.29$ GeV~\cite{Yao:2006px}.

\begin{figure}[t]
\includegraphics[width=40ex]{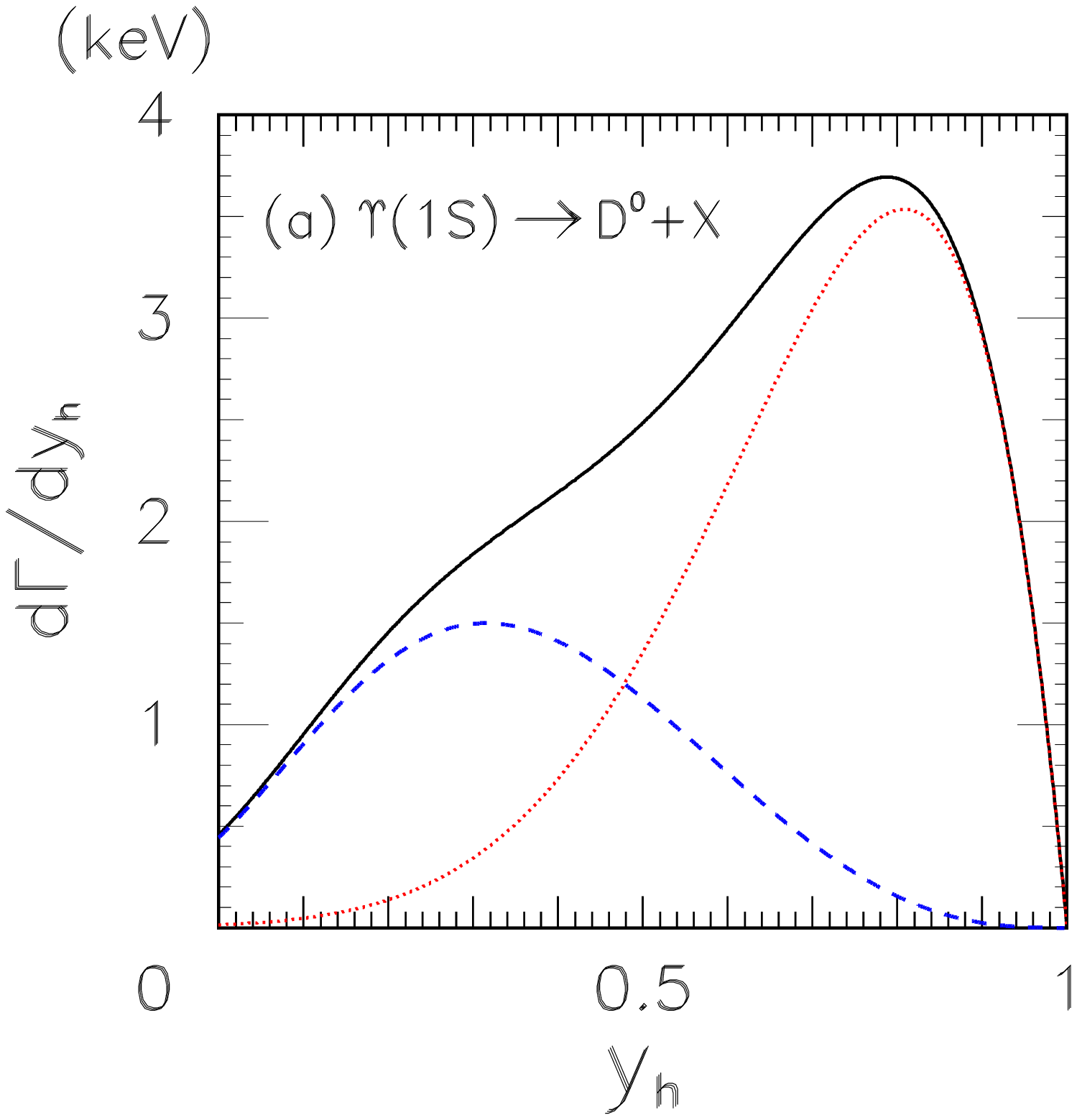}\hspace{7ex}
\includegraphics[width=40ex]{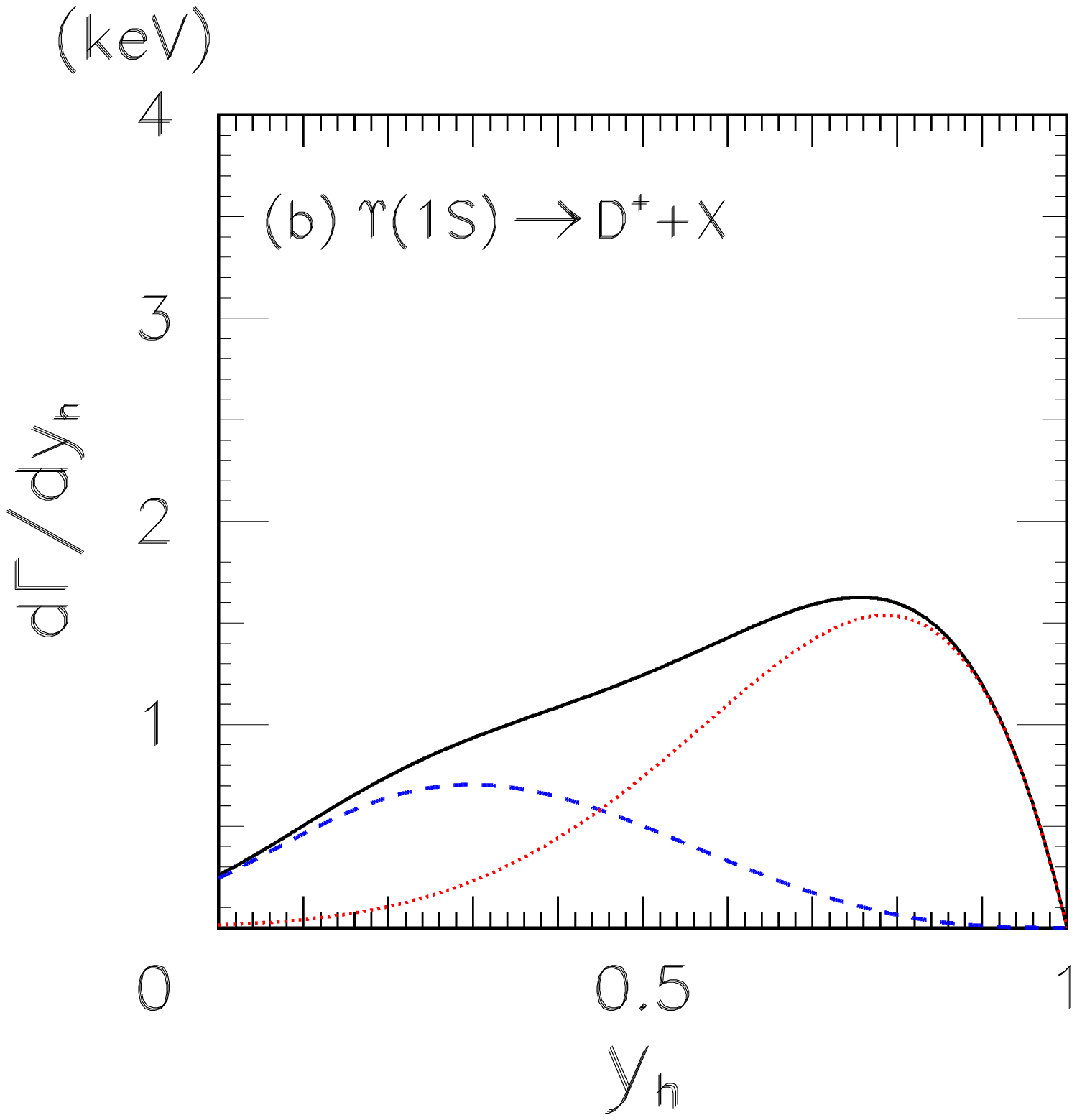}
\\
\includegraphics[width=40ex]{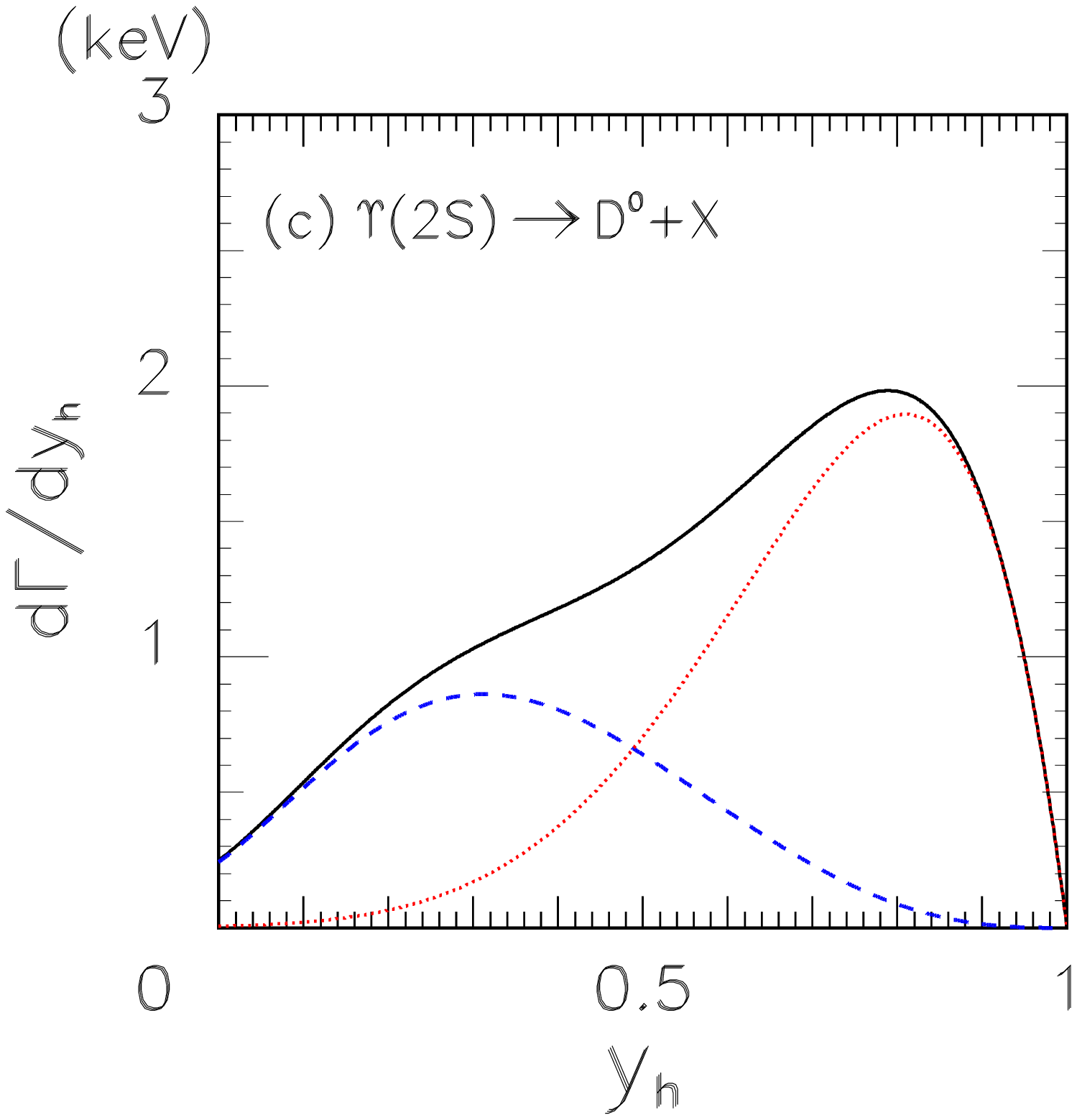}\hspace{7ex}
\includegraphics[width=40ex]{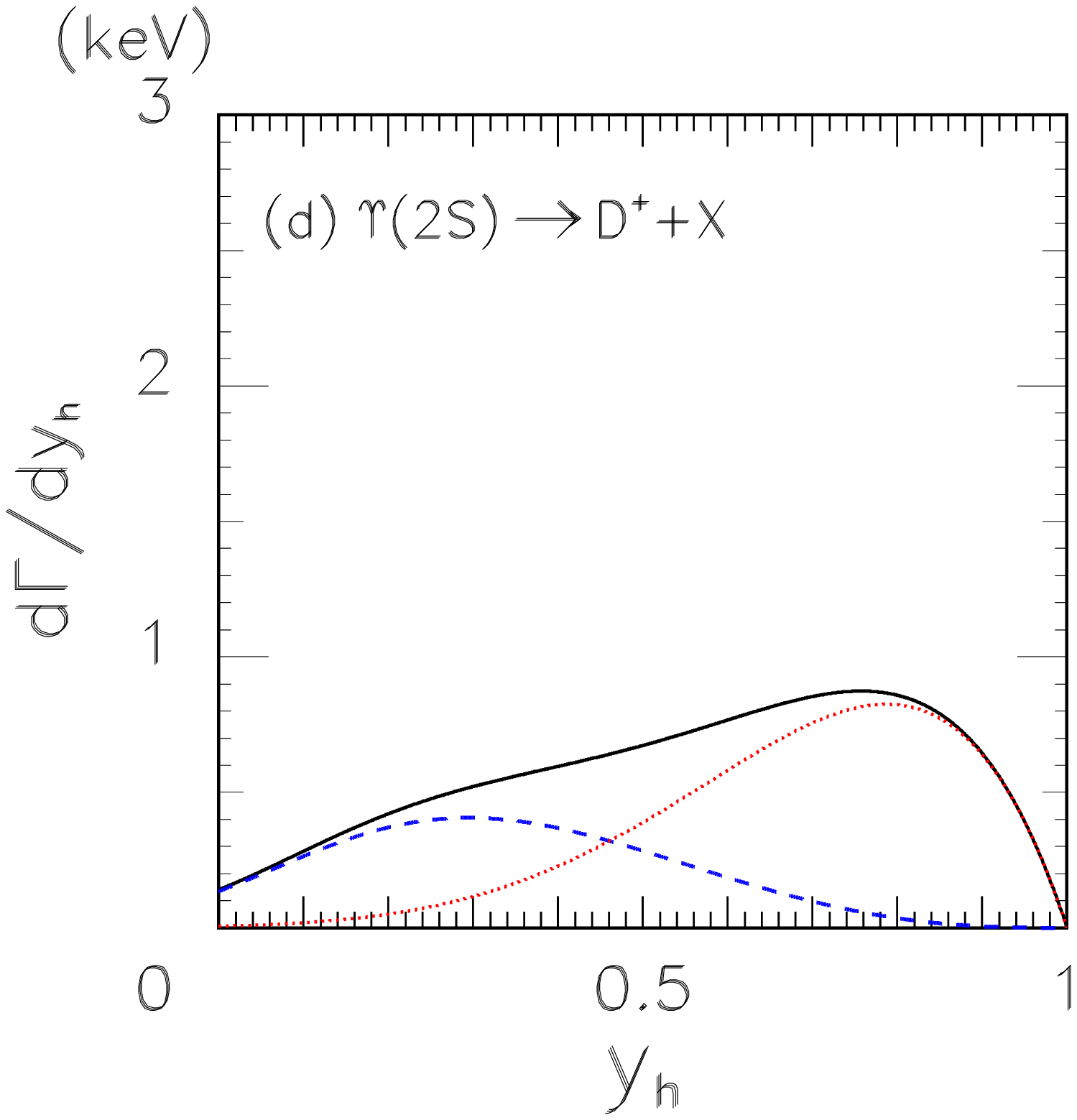}
\\
\includegraphics[width=40ex]{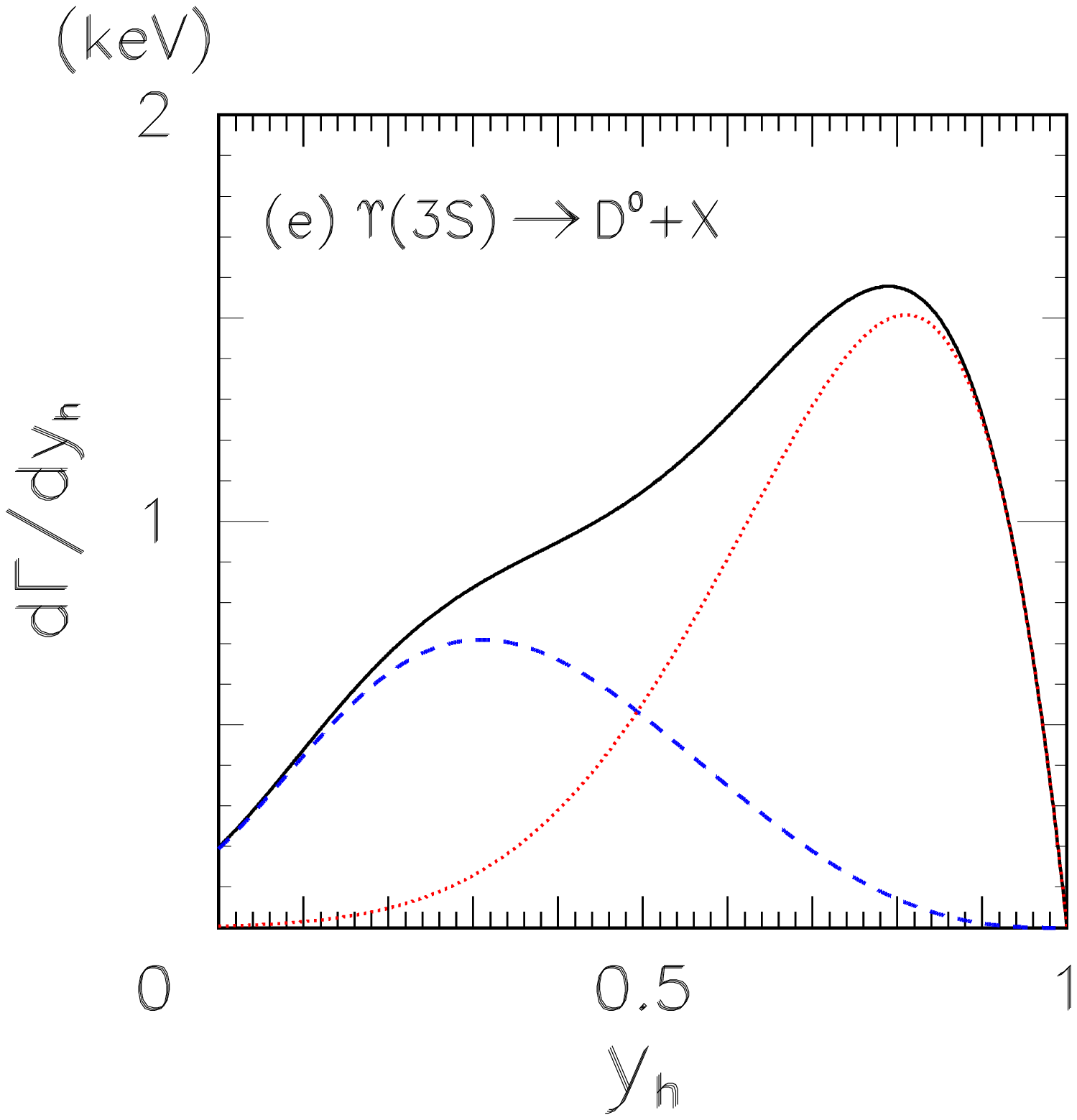}\hspace{7ex}
\includegraphics[width=40ex]{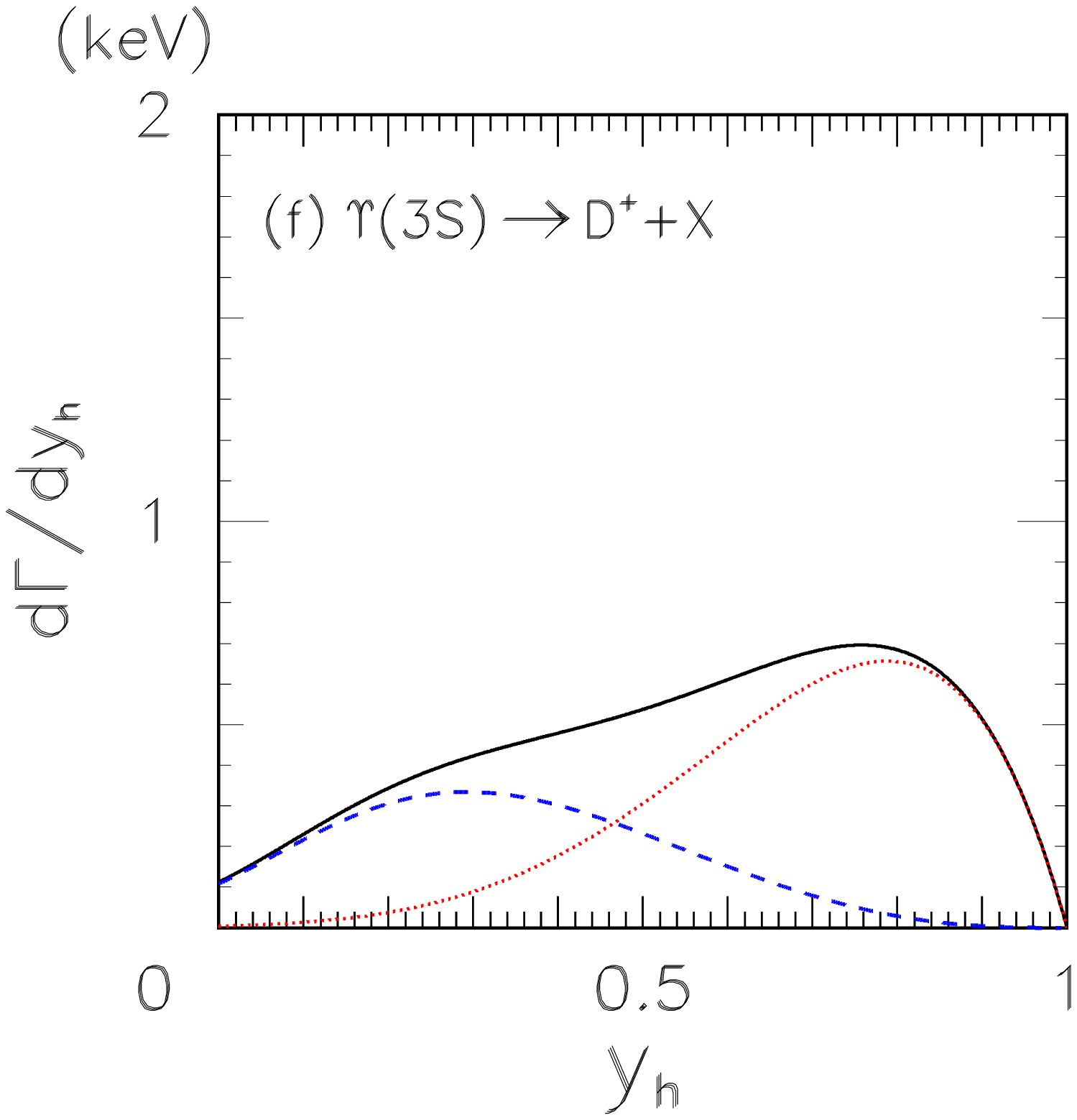}
\\[-2.0ex]
\caption{
Distributions of the scaled momentum $y_h$ in $\Upsilon(nS)\to h+X$ 
for $h=D^0$ (left column) and $D^+$ (right column) in units of keV.
KLP parameterization is used for the charm fragmentation 
functions~\cite{Seuster:2005tr}. In each figure, solid, dashed, and dotted
curves represent the total, QCD ($c/g^*$), and virtual-photon ($c/\gamma^*$) 
contributions, respectively.
}
\label{fig3}%
\end{figure}

\begin{figure}[t]
\includegraphics[width=40ex]{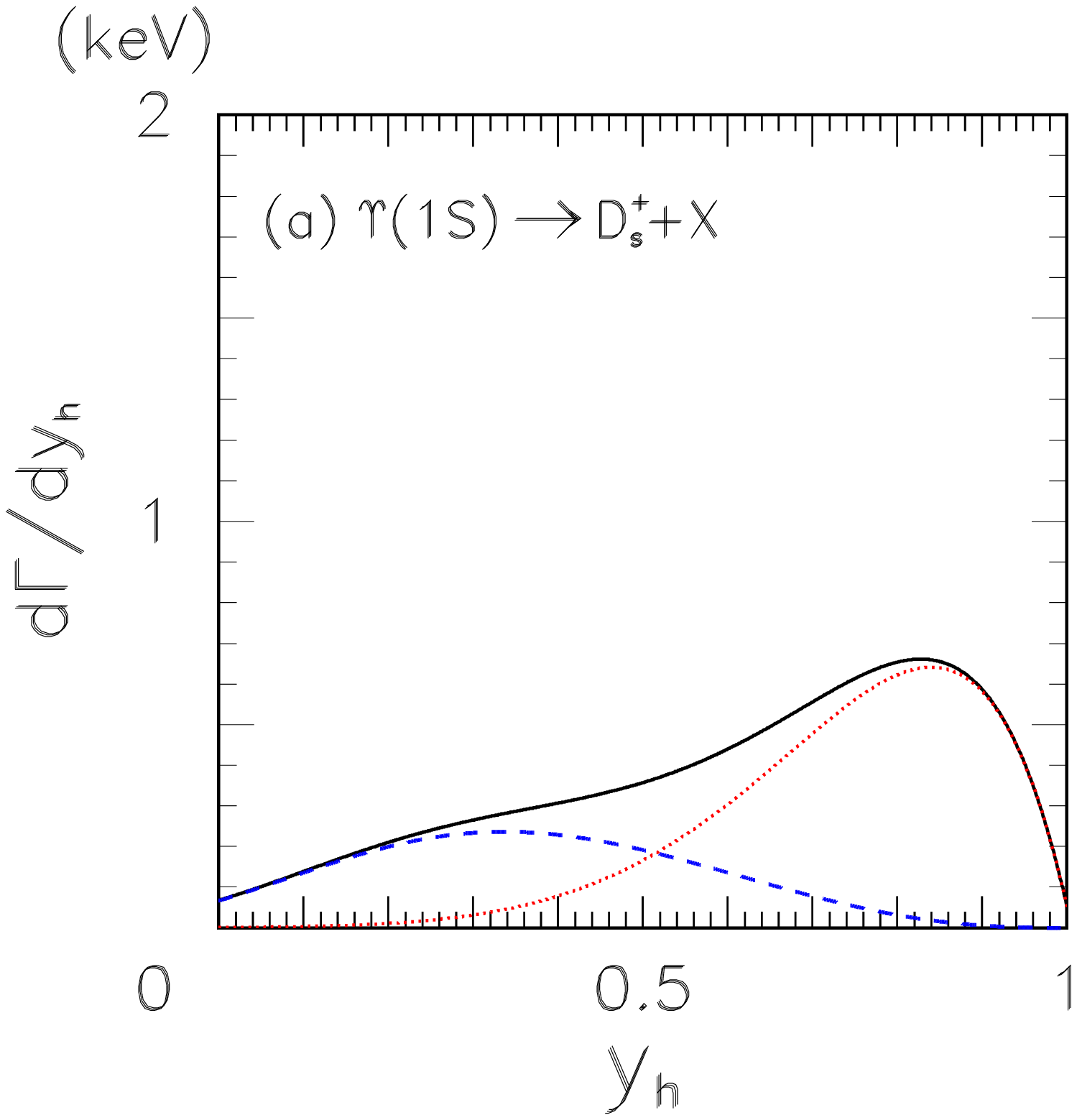}\hspace{7ex}
\includegraphics[width=40ex]{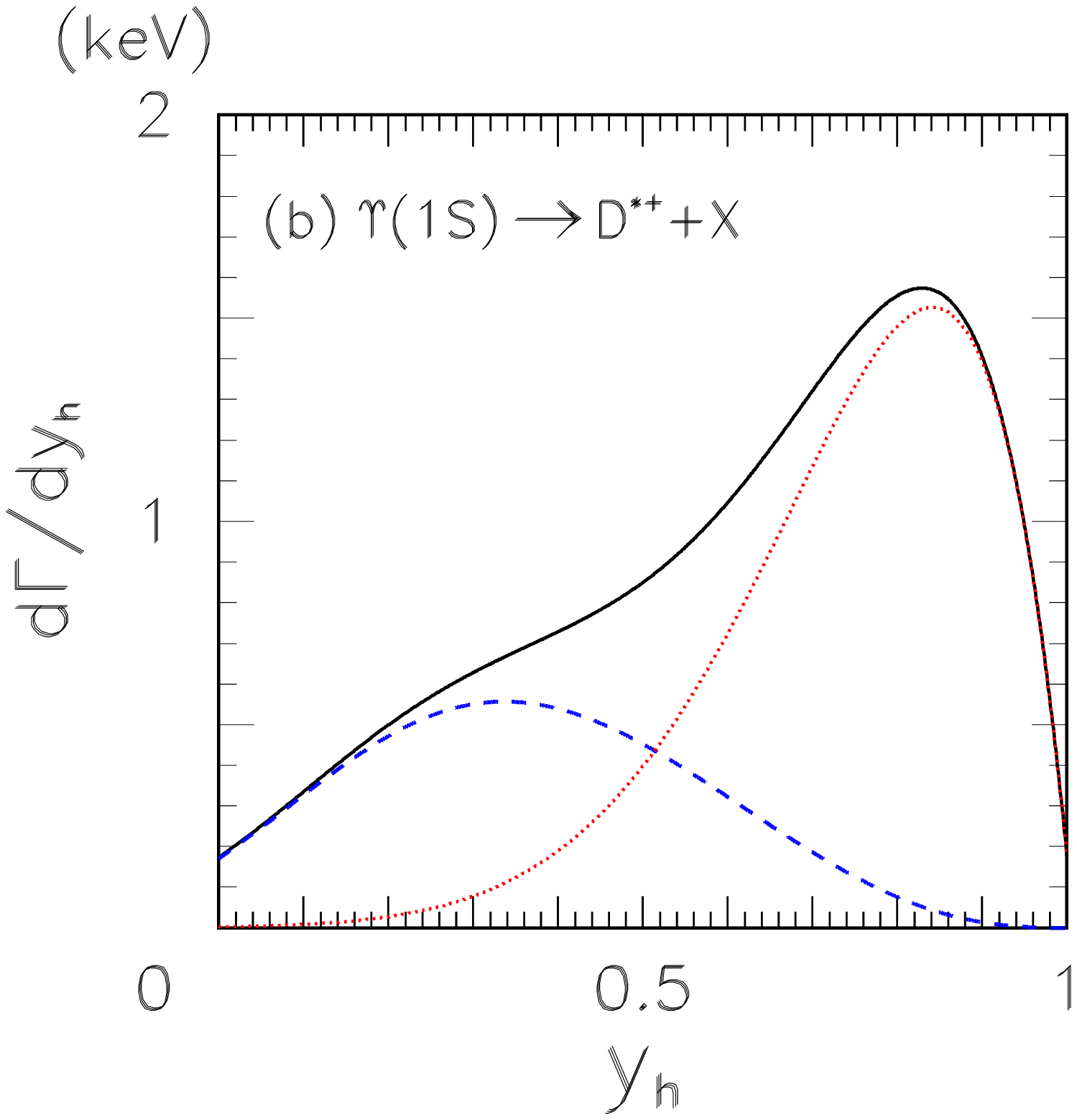}
\\
\includegraphics[width=40ex]{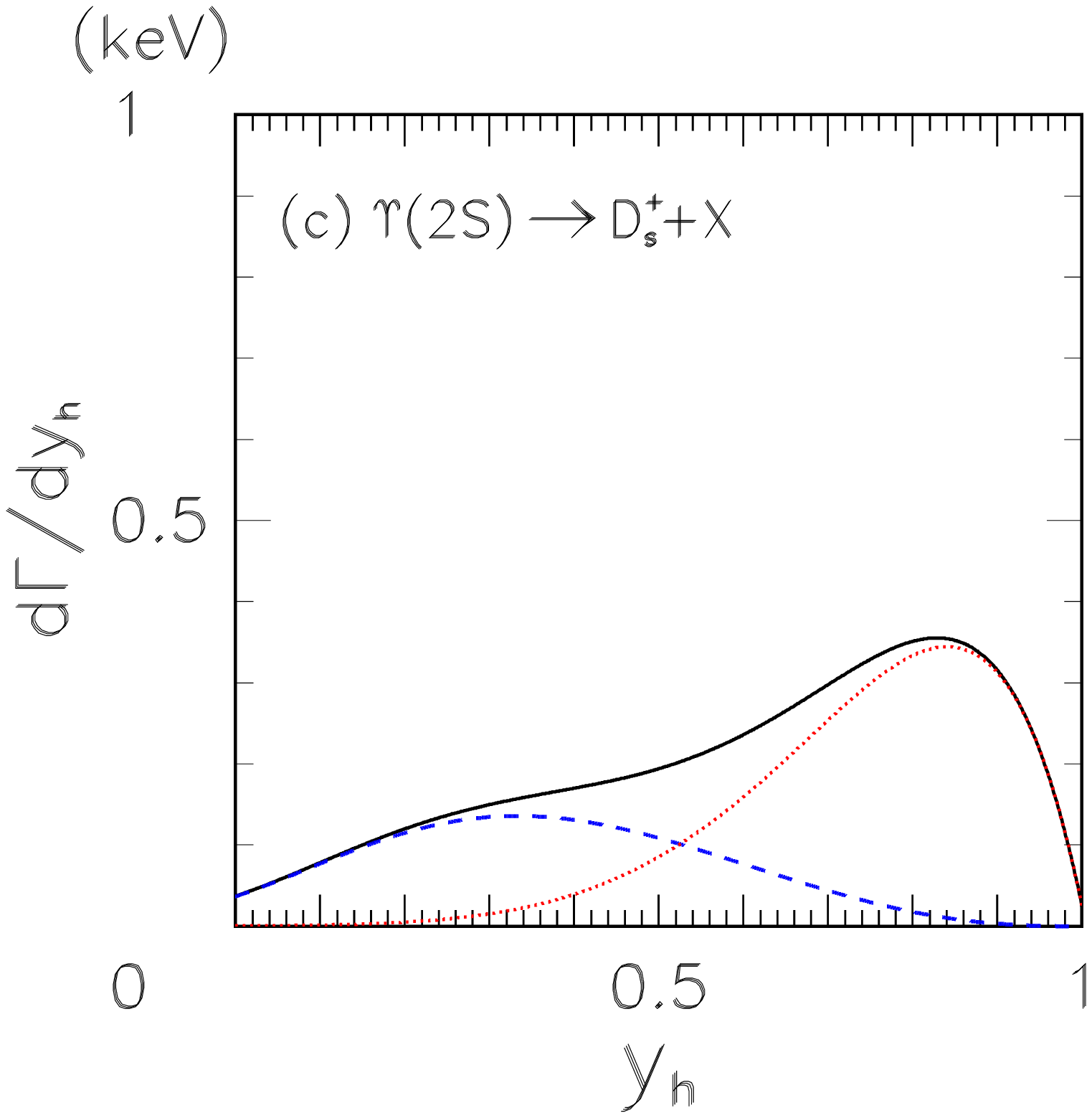}\hspace{7ex}
\includegraphics[width=40ex]{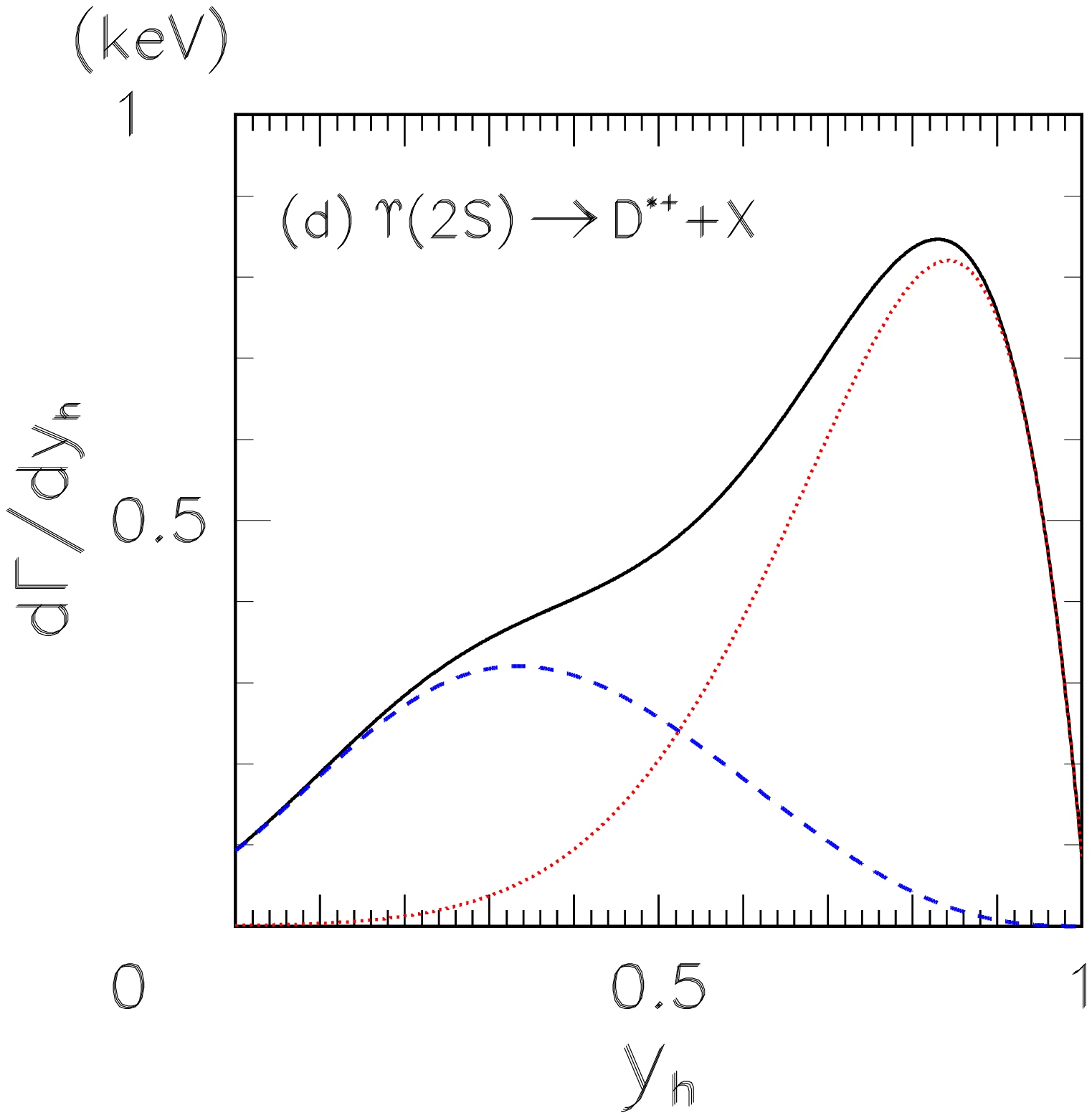}
\\
\includegraphics[width=40ex]{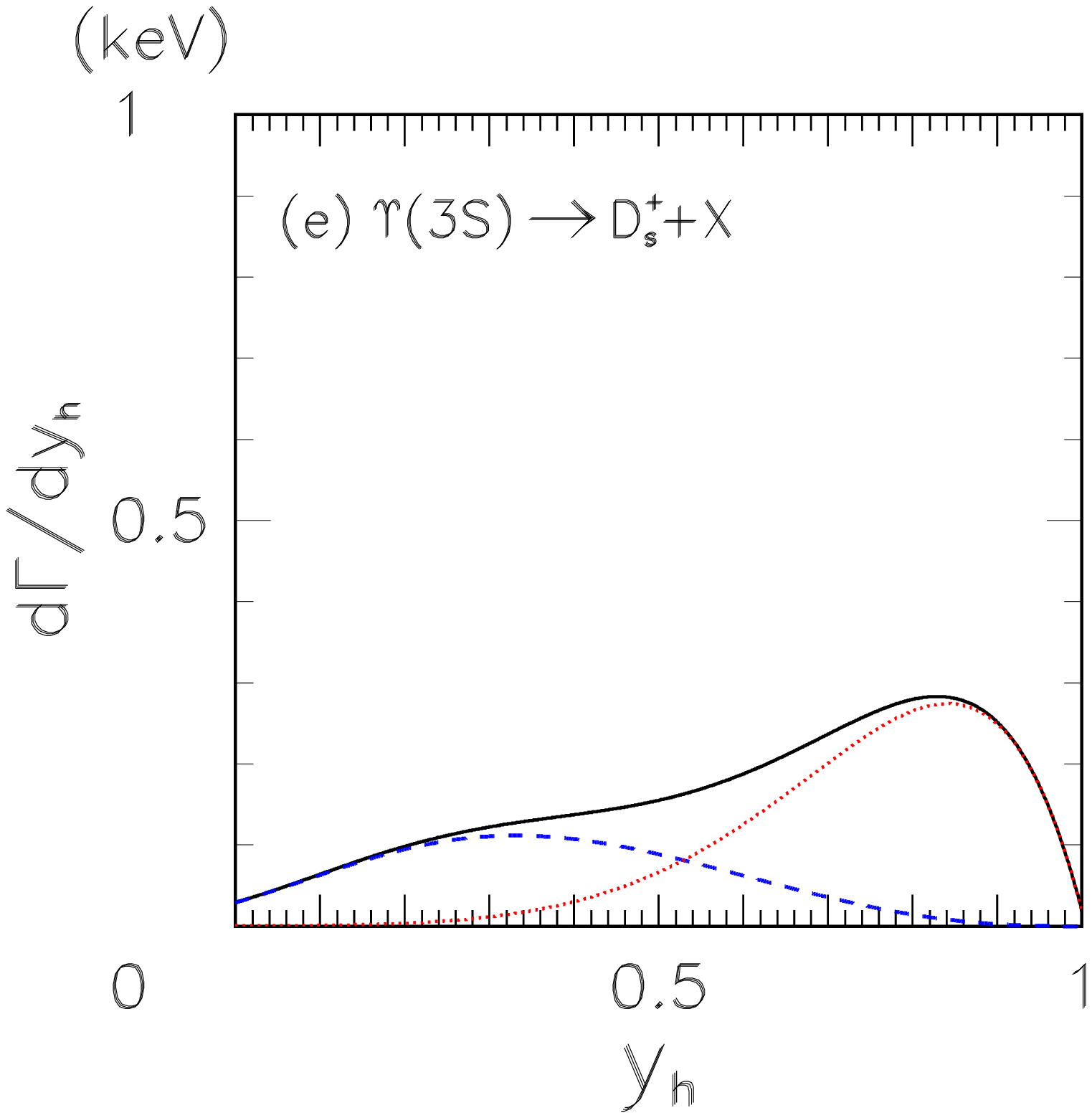}\hspace{7ex}
\includegraphics[width=40ex]{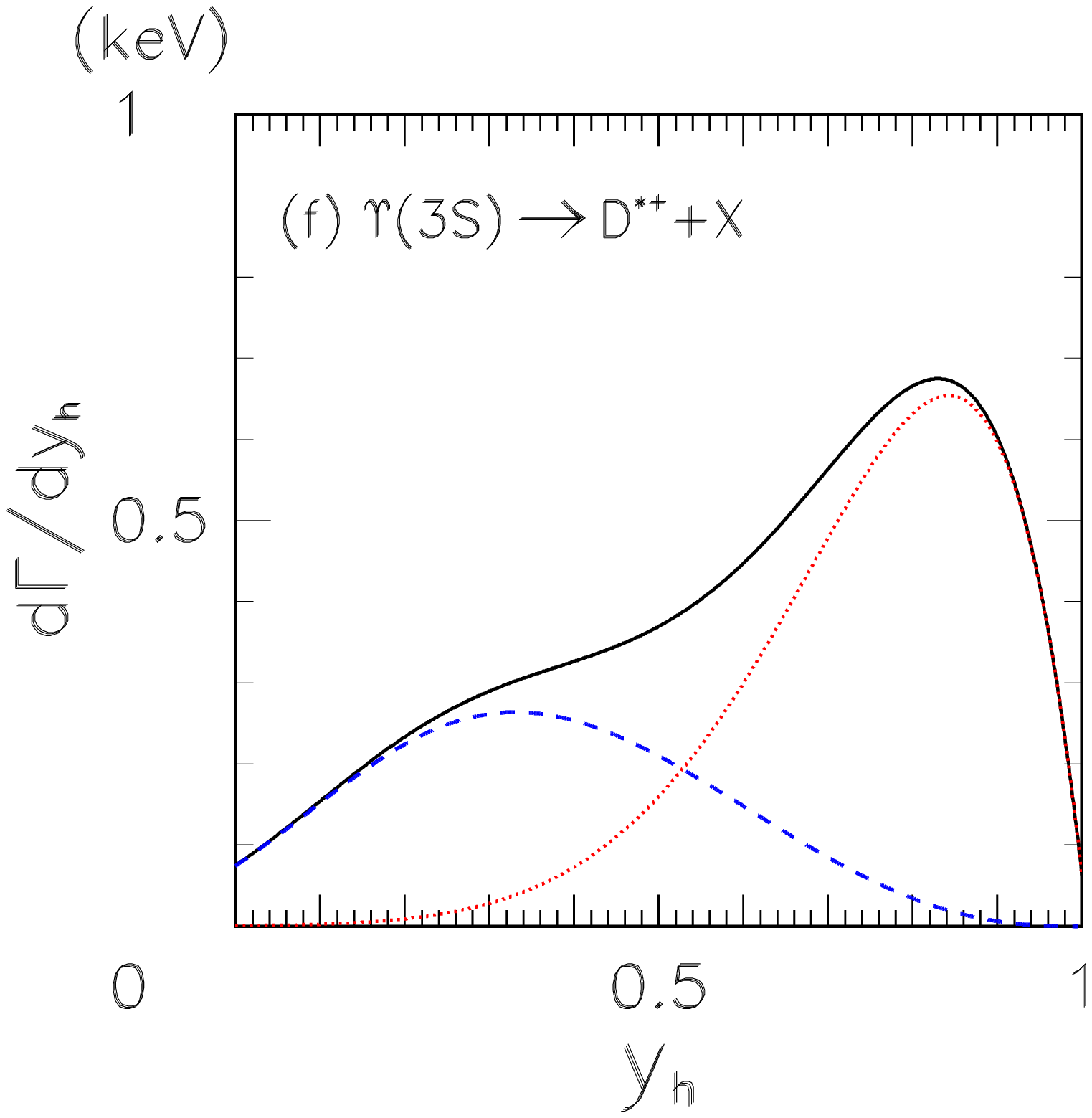}
\\[-2.0ex]
\caption{
Distributions of the scaled momentum $y_h$ in $\Upsilon(nS)\to h+X$
for $h=D^{+}_s$ (left column) and $D^{*+}$ (right column) in units of keV.
KLP parameterization is used for the charm fragmentation
functions~\cite{Seuster:2005tr}. In each figure, solid, dashed, and dotted
curves represent the total, QCD ($c/g^*$), and virtual-photon ($c/\gamma^*$)
contributions, respectively.
}
\label{fig4}%
\end{figure}

\begin{figure}[t]
\includegraphics[width=40ex]{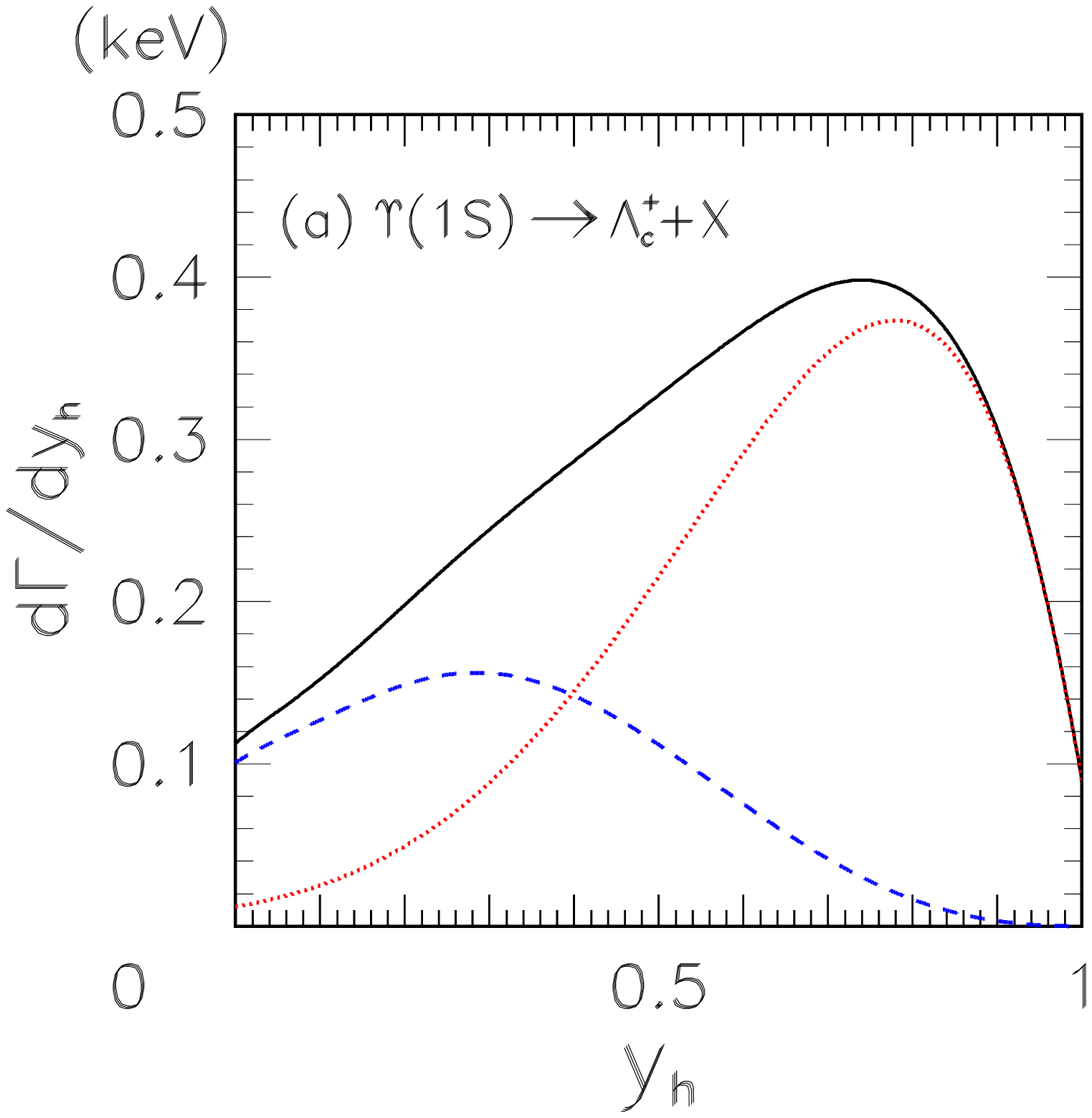}
\\
\includegraphics[width=40ex]{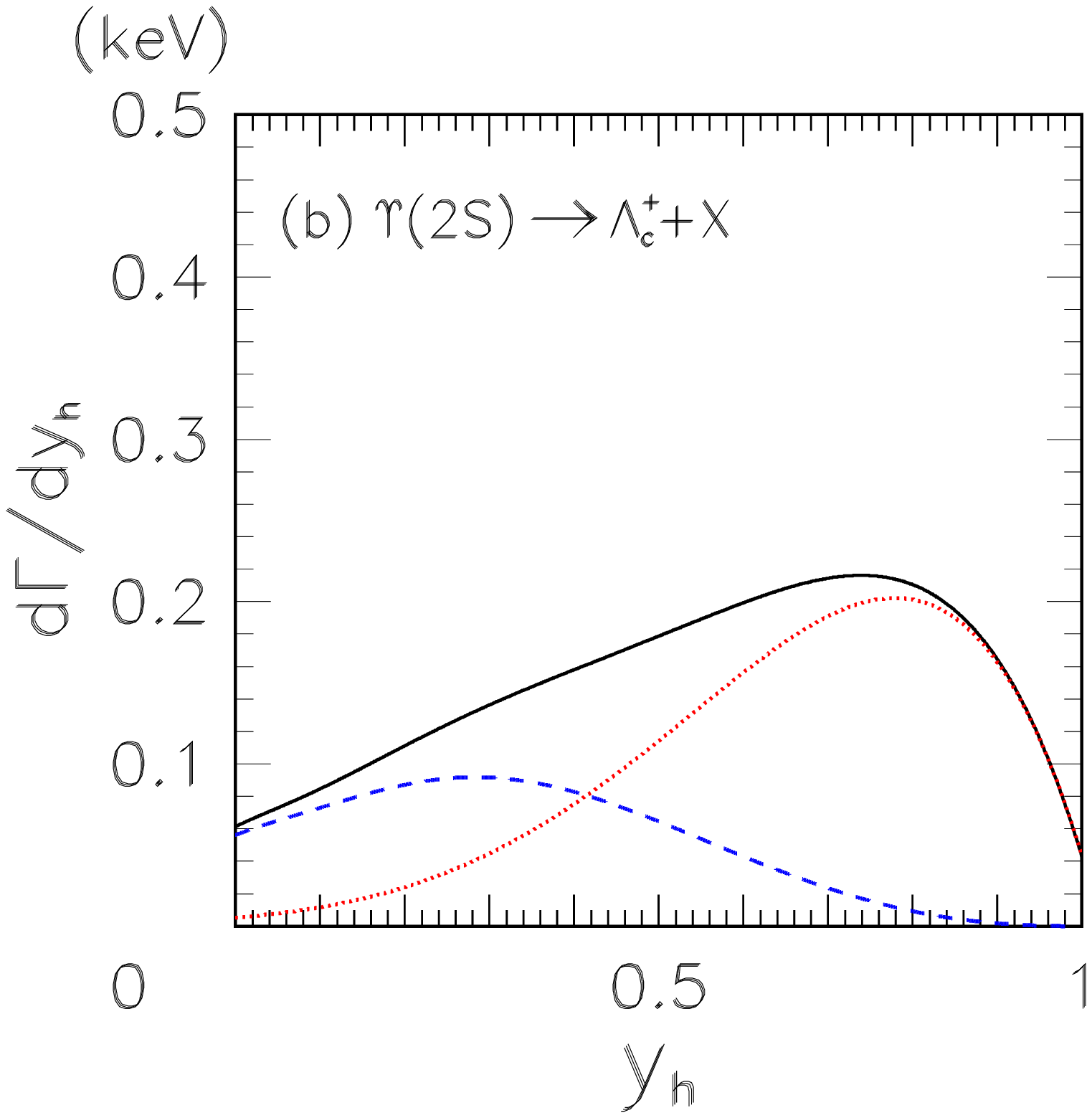}
\\
\includegraphics[width=40ex]{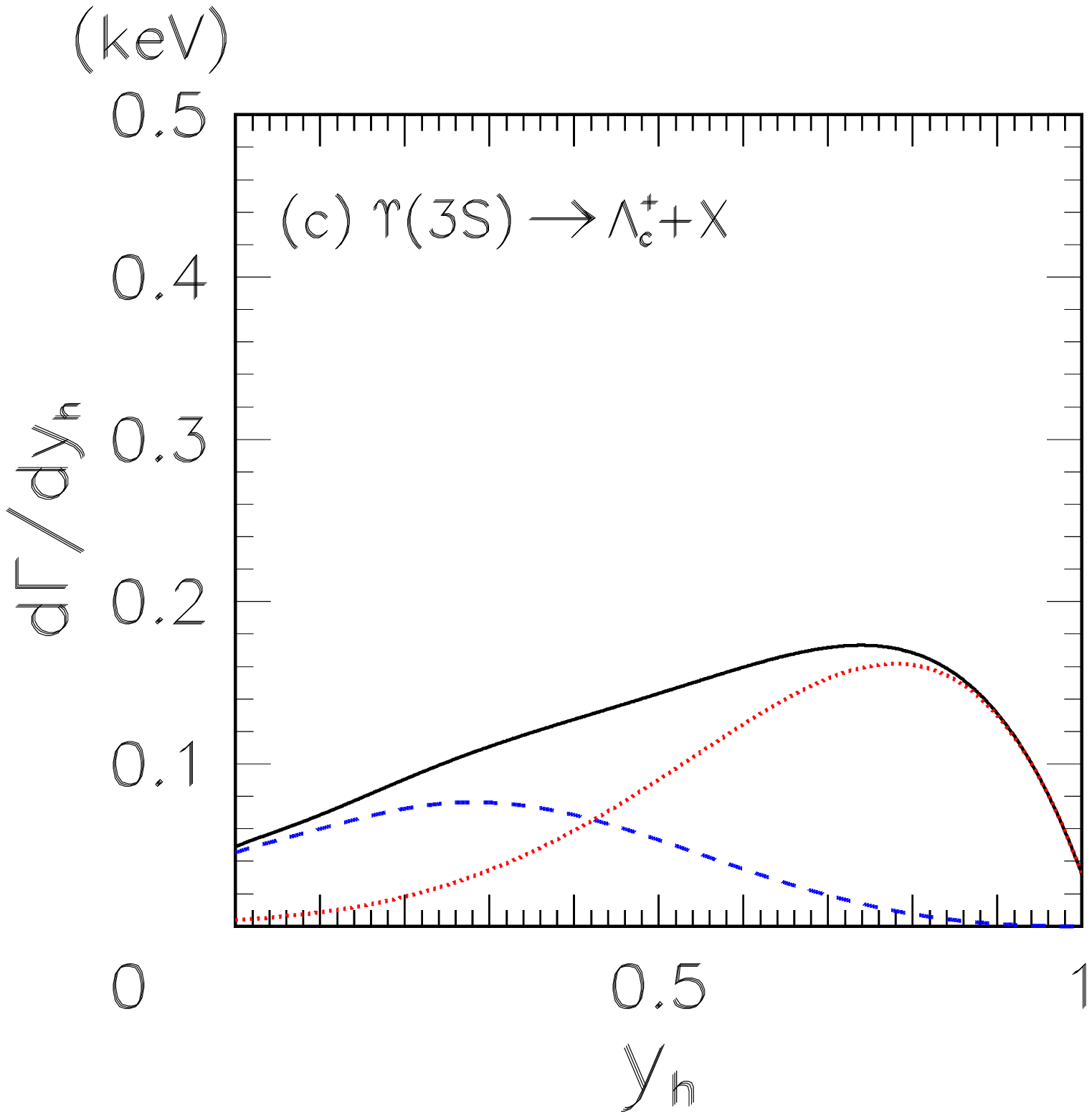}
\\[-2.0ex]
\caption{
Distributions of the scaled momentum $y_h$ in 
$\Upsilon(nS)\to \Lambda_c^++X$ in units of keV.
KLP parameterization is used for the charm fragmentation
functions~\cite{Seuster:2005tr}. In each figure, solid, dashed, and dotted
curves represent the total, QCD ($c/g^*$), and virtual-photon ($c/\gamma^*$)
contributions, respectively.
}
\label{fig5}%
\end{figure}

In Fig.~\ref{fig3}, we show the scaled momentum $y_h$ distributions
of the charmed hadrons in the decays $\Upsilon(nS)\to h+X$, 
where $h=D^0$ (left column) and $D^+$ (right column) for radial quantum
numbers $n=1$, 2, and 3. The distributions for $\Upsilon(nS)\to h+X$, 
where $h=D^+_s$ (left column) and $D^{*+}$ (right column) are illustrated
in Fig.~\ref{fig4} and those for the $\Lambda_c^+$ baryon in Fig.~\ref{fig5}.
In Figs.~\ref{fig3}, \ref{fig4}, and \ref{fig5}, the dotted and dashed curves 
represent the virtual-photon and QCD contributions, respectively. 
The solid lines are the
sums of the two contributions. In contrast to the charm-quark
momentum distribution, which has a virtual-photon contribution only 
at the end point
$y_1=1$, the virtual-photon contribution for a charmed hadron is 
smeared out to the region $y_h<1$. 
The peaks of the virtual-photon, QCD, and total contributions are placed in the
ranges $0.78<y_h<0.84$, $0.28<y_h<0.34$, and $0.75<y_h<0.83$, respectively. 
The softening of the QCD contribution can be seen by comparing the
QCD contribution to the charmed-hadron momentum distributions in Figs.~\ref{fig3},
\ref{fig4}, and \ref{fig5} with the charm-quark momentum distributions in 
Fig.~\ref{fig2}. The softening of the virtual-photon contribution is even
more dramatic, transforming a delta function at $y_1=1$ into the
virtual-photon contributions in Figs.~\ref{fig3}, \ref{fig4}, and 
\ref{fig5}.  Hadronization significantly softens the momentum spectrum 
of charmed hadrons.

The charmed-hadron production rates in $\Upsilon$ decay are obtained 
by integrating over $y_h$ in Eq.~(\ref{dGamD}) for each hadron.
The rates are the same as the area under the solid lines 
in Figs.~\ref{fig3}, \ref{fig4}, and \ref{fig5}.
The production rates of $D^0$, $D^+$, $D_s^+$, $D^{* 0}$,
and $\Lambda_c^+$ in $\Upsilon(1S)$ decay are 
$2.25$ keV, $1.06$ keV, $0.37$ keV, $0.87$ keV, 
and $0.28$ keV, respectively,
where the contributions of the virtual-photon processes
amount to about $65$--$\,69\%$ of the production rates.
In $\Upsilon(2S)$ decay, we get
$1.23$ keV, $0.58$ keV, $0.20$ keV, $0.47$ keV, and $0.15$ keV
for each hadron production rate, respectively
while in $\Upsilon(3S)$ decay, 
$0.98$ keV, $0.46$ keV, $0.16$ keV, $0.38$ keV, and $0.12$ keV, respectively.
For the decays of the $2S$ state into charmed hadrons
the virtual-photon contributions add up to about $63$--$\,67 \%$
while for those of the $3S$ state to about $63$--$\,66 \%$.
The fractions of the virtual-photon contributions 
decrease up to about $63\%$ 
as the radial quantum number $n$ of the $\Upsilon(nS)$ increases.
If we compare the momentum distributions of charmed
hadrons with that of the charm quark, we could observe that
hadronization softens the distributions. For the QCD contributions,
whose charm-quark distribution reaches lower end point, the vertical-axis 
intercepts for the charmed-hadron momentum distributions
significantly shift to the positive direction. This results in loss of
probability. Because the loss is minor for the virtual-photon
contributions, whose charm-quark distributions are concentrated at
$y_1=1$, the fractions of the virtual-photon contributions in
the charmed-hadron production rates are greater than that of the
charm quark production rate.
As is discussed in Ref.~\cite{Bodwin:2007zf}, this change in
normalization is of order $r$, which is at the level of the
error in the fragmentation approximation itself,
which is derived from QCD by
neglecting corrections on the order of the square of the quark mass
divided by the hard-scattering momentum \cite{Collins:1989gx}.

As we show in Figs.~\ref{fig3},
the production rate of $D^0$ is more than twice as large as
that of $D^+$.
This reflects the fact that the $D^0$ may be produced from the decays of
$D^{* 0}$ and $D^{*+}$ 
while $D^+$ has the contribution only from the feeddown from the decay
of $D^{*+}$ besides direct production from the charm quark.
As we show in Fig.~\ref{fig5}, the shape of the scaled momentum distribution
of the $\Lambda_c^+$ is broader than that of the other charmed hadrons.
Because ${\Lambda_c^+}$ is the heaviest among the charmed hadrons 
that we consider,
the smearing effects from hadronization 
is most significant in $\Lambda_c^+$ production.

There are uncertainties in the fragmentation function. 
Part of the uncertainties can be estimated by
comparing the results shown above with those obtained by using
different parameterizations for the fragmentation functions. 
We also carried out the same calculations by using the Collins-Spiller (CS) 
fragmentation functions~\cite{Collins:1984ms} whose parameters are 
determined in Ref.~\cite{Seuster:2005tr}. With
the CS fragmentation functions, the heights of the peaks are smaller
than those from the KLP fragmentation functions 
by about 8\%. This difference is
small enough to be within our theoretical uncertainties of about
20\%.
Another important source of the uncertainties may come from
the Monte Carlo. The estimate of the uncertainties is out of the
scope of this work.


\section{Summary}
\label{summary}
We have studied inclusive charm production in the decay of 
the spin-triplet bottomonium state $\Upsilon(nS)$ 
based on the color-singlet mechanism of the NRQCD factorization
formalism. The branching fractions of the $\Upsilon(nS)$ 
into charm are predicted to be
$(7.5\pm 1.4)\%$,
$(6.9\pm 1.4)\%$, and
$(8.7\pm 1.7)\%$ for $n=1$, $2$, and $3$, respectively.
The dominant contribution comes from the virtual-photon process
$b\bar{b}_1({}^3S_1)\to\gamma^* \to c\bar{c}$, which contributes
about 60\% of the partial
width for inclusive charm production in each $\Upsilon(nS)$ decay.
The remaining portion of the partial width comes from
the QCD contribution, which is dominated by 
$b\bar{b}_1({}^3S_1)\to ggg^*$ followed by $g^*\to c\bar{c}$.
The $b\bar{b}_1(^3S_1)\to c\bar{c}g \gamma$ diagrams contribute
only about $0.7\%$ of the total widths of $\Upsilon(nS)$ into charm. 

We have also presented the momentum distribution of the charm quark
in the $\Upsilon(nS)$ decays. The virtual-photon contribution is localized at
the end point ($y_1=1$). The QCD contributions, which have
four-body final states, are broad and have peaks near $y_1=0.5$.
By convolving the charm momentum distribution with the fragmentation
function $D_{c\to h}(z)$ that have been fit to the $e^+e^-$ annihilation
data by the Belle Collaboration, we have computed the momentum distribution
of charmed hadrons $h=D^0$, $D^+$, $D_s^+$, $D^{*+}$, and $\Lambda_c^+$
produced in $\Upsilon(nS)\to h+X$.
The resulting momentum distributions of the charmed hadrons are
significantly softer than that of the charm quark.
The large virtual-photon contribution, which is localized at the end
point ($y_1=1$) for the charm quark, is smeared out into a distribution
with a peak near $y_h\approx 0.8$.
The order-$\alpha_s$ corrections to the virtual-photon
contributions may enhance the charm-quark momentum distribution
in the range $y_1<1$. The cancellation of the infrared divergence
at that order may produce a singular distribution at the end point,
which is similar to that of the $P$-wave decays.

In a recent analysis being carried out by the CLEO Collaboration,
as a part of measurement, the virtual-photon contribution is subtracted from
the data. In order to facilitate comparison with the experimental data,
we have provided all the predictions separately for the QCD, virtual-photon,
and total contributions. Comparison with the experimental results will test
the leading-order approximations employed in this work.

\begin{acknowledgments}
We thank Roy Briere for suggesting this problem and for useful
discussions. We thank Bostjan Golob, Soeren Lange, and Rolf Seuster
for their kind explanation of the analysis given in 
Ref.~\cite{Seuster:2005tr}. We also express our gratitude to Geoff~Bodwin
and Eric~Braaten for valuable comments and suggestions.
We also appreciate Hee Sok Chung for helping us with calculating
the NRQCD matrix elements.
JL thanks the High Energy Physics Theory Group at the Ohio State
University for its hospitality while this work was being completed.
This work was supported in part by the BK21 program of Ministry of
Education, Korea.
D.K., J.L., and C.Y. were supported by the Korea Research Foundation
under grants KRF-2004-015-C00092, KRF-2006-311-C00020, and 
KRF-2005-075-C00008, respectively.
J.L. was also supported by a Korea University grant.
T.K. was supported by
the Basic Research Program of the Korea Science and Engineering
Foundation (KOSEF) under grant No.~R01-2005-000-10089-0.

\end{acknowledgments}
\appendix
\section{The four-body phase space\label{app}}
The four-body phase space $d \Phi_4$ is defined by
\begin{equation}
d\Phi_4
=(2\pi)^{4}\delta^{(4)}\left(P-\sum_{i=1}^4 p_i\right)
\prod_{i=1}^4
\frac{d^{3}{{p_i}}}{(2\pi)^{3}2E_i},
\label{PS4-def-app}%
\end{equation}
where $E_i$ and $p_i$ are the energy and momentum 
of the particle $i$ in the final state.
For $\Upsilon(P)\to c(p_1)\bar{c}(p_2) g(p_3)g(p_4)$, $p_1^2=p_2^2=m_c^2$, 
$p_3^2=p_4^2=0$, and $\sqrt{P^2}=2E_b$. 
Since we are interested 
in the momentum distribution of the charm quark in the $P$-rest frame, 
we evaluate
$d\Phi_4$ leaving the three-momentum $\bm{p}_1$ unintegrated:
\begin{equation}
d{\Phi_4} =
\frac{d^3{p}_1}{(2\pi)^3 2E_1} d{\Phi}_3(X\to p_2+p_3+p_4),
\label{dphi4-1}%
\end{equation}
where $X=P-p_1$.
The three-body phase space $d{\Phi}_3(X\to p_2+p_3+p_4)$
can be expresses as a chain of two-body phase spaces:
\begin{equation}
d{\Phi}_3(X\to p_2+p_3+p_4)=
d{\Phi}_2(X\to p_2+Y) \,\frac{dm_Y^2}{2\pi} \,d{\Phi}_2(Y\to p_3+p_4),
\label{dphi3}%
\end{equation}
where $Y=p_3+p_4$ and $m_Y$ is the invariant mass of $Y$.
When a squared amplitude is summed over spin states of 
all the particles in both initial and final states,
the squared amplitude becomes independent of the solid angle of $\bm{p}_1$. 
Integrating over the solid angle of $\bm{p}_1$,
substituting Eq.~(\ref{dphi3}) into Eq.~(\ref{dphi4-1}), and
expressing the two-body phase spaces $d{\Phi}_2(X\to p_2+Y)$ and
$d{\Phi}_2(Y\to p_3+p_4)$ in the $X$ and $Y$ rest frames,
respectively, we find that
\begin{equation}
d{\Phi}_4=
\frac{|\bm{p}_1| |\bm{p}_2^*| |\bm{p}_3^*|}{2^{10}\pi^7 m_X}
\,dE_1 dm_Y d\Omega_2^* d\Omega_3^*,
\label{dphi4}%
\end{equation}
where $m_X=(4 E_b^2 -4 E_b E_1 + m_c^2)^{1/2}$ is the invariant mass of $X$.
The ranges of the integration variables $E_1$ and $m_Y$ are given by
\begin{subequations}
\begin{eqnarray}
\label{E1-range}%
m_c\leq &E_1& \leq E_b,
\\
\label{mY-range}%
0 \leq &m_Y& \leq m_X-m_c.
\end{eqnarray}
\label{range-em}%
\end{subequations}
In Eq.~(\ref{dphi4}), 
$|\bm{p_1}|$ and $E_1$ are the absolute value of the charm-quark momentum 
and energy
in the $P$-rest frame while
$\bm{p}_2^*$ $(\bm{p}_3^*)$ and
$d\Omega_2^*$ $(d\Omega_3^*)$ are the three-momentum
and the solid-angle element of the $\bar{c}$ $(g)$
in the $X$ $(Y)$-rest frame, respectively.
Explicit components of the four-vectors $p_1$, $p_2^*$ and $p_3^*$ are
\begin{subequations}
\begin{eqnarray}
{p}_1&=&(E_1,0,0,|\bm{p}_1|),
\\
{p}_2^*&=&(E_2^*,|\bm{p}_2^*|\sin\theta_2^*\cos\phi_2^*,
             |\bm{p}_2^*|\sin\theta_2^*\sin\phi_2^*,
             |\bm{p}_2^*|\cos\theta_2^*),
\\
{p}_3^*&=&
           (|\bm{p}_3^*|,|\bm{p}_3^*|\sin\theta_3^*\cos\phi_3^*,
            |\bm{p}_3^*| \sin\theta_3^*\sin\phi_3^*,
            |\bm{p}_3^*|\cos\theta_3^*),
\end{eqnarray}
\label{momcomp}%
\end{subequations}
where 
($\theta_2^\ast$, $\phi_2^\ast$) and ($\theta_3^\ast$, $\phi_3^\ast$)
are the polar and azimuthal angles of $\bm{p}_2^\ast$ 
and $\bm{p}_3^\ast$ in the $X$-rest and $Y$-rest frame, respectively.
In Eq.~(\ref{momcomp}), 
\begin{subequations}
\begin{eqnarray}
E_2^\ast &=& \frac{m_X^2+m_c^2-m_Y^2}{2m_X}, \\
|\bm{p}_1| &=& (E_1^2 - m_c^2)^{1/2}, \\
|\bm{p}_2^\ast| &=& \frac{1}{2 m_X} \lambda^{1/2}(m_X^2,m_Y^2,m_c^2), \\
|\bm{p}_3^\ast| &=& \frac{m_Y}{2},
\end{eqnarray}
\end{subequations}
where $\lambda(a,b,c)=a^2+b^2+c^2-2 a b - 2 b c - 2 c a$.
Note that in Eq.~(\ref{momcomp}) $p_2^\ast$ and $p_3^\ast$ are 
given in the $X$-rest and $Y$-rest frame
while in order to evaluate Eq.~(\ref{dC}) it is convenient to express
them in the $P$-rest frame.
We introduce the boost matrix ${\Lambda^\mu}_\nu$
transforming an arbitrary vector $k^\ast=(\sqrt{k^2},\bm{0})$ 
into $k=(k^0,\bm{k})$:
\begin{subequations}
\begin{eqnarray}
k^\mu&=&{\Lambda^\mu}_\nu k^{\ast\nu}, \\
{\Lambda^0}_0&=&\frac{k^0}{\sqrt{k^2}}, \\
~{\Lambda^0}_i&=&{\Lambda^i}_0=\frac{k^i}{\sqrt{k^2}}, \\
{\Lambda^i}_j&=&\delta^{ij} 
+ \frac{k^0-\sqrt{k^2}}{\sqrt{k^2}}\frac{k^i k^j}{|\bm{k}|^2},
\end{eqnarray}
\label{boost}%
\end{subequations}
where $i$, $j = 1$, 2, 3.
The explicit components of $p_2$ are obtained 
by boosting $p_2^\ast$
from the $X$-rest frame to the $P$-rest frame,
where ${\Lambda^\mu}_\nu$ is determined by 
substituting $X$ into $k$ in Eq.~(\ref{boost}).
To obtain $p_3$, we need two steps.
First, we boost $p_3^\ast$ from the $Y$-rest frame to the $X$-rest frame
where the boost matrix is given by 
replacing $k$ in Eq.~(\ref{boost}) by $Y$.
The components of $Y$ in the $X$-rest frame are easily obtained by using
$Y=X-p_2$ valid in any frame.
Let the obtained four-vector be $p_3^X$.
It is easy to obtain $p_3$ by boosting $p_3^X$ from the $X$-rest frame
to the $P$-rest frame.
Now we can express all the momenta in the $P$-rest frame so that
the Lorentz scalars in Eq.~(\ref{dC}) can be represented in the
$P$-rest frame.

It is convenient to introduce dimensionless variables
$x_1$ and $r_Y$ defined by
\begin{subequations}
\begin{eqnarray}
x_1 &=& E_1/E_b, \\
r_Y &=& m_Y/E_b,
\end{eqnarray}
\label{xi-def}%
\end{subequations}
where the ranges of the variables are
\begin{subequations}
\begin{eqnarray}
\label{x1-range}%
\sqrt{r_c}\leq &x_1& \leq1
,
\\
\label{xY-range}%
0 \leq &r_Y& \leq \sqrt{4 - 4x_1 + r_c}-\sqrt{r_c}.
\end{eqnarray}
\label{range}%
\end{subequations}
$r_c$ is the square of the ratio of the charm-quark mass and $E_b$,
$r_c=m_c^2/E_b^2$.
The energy and momenta 
$E_2^\ast$, $|\bm{p}_1|,\,|\bm{p}_2^*|$, and $|\bm{p}_3^*|$
are expressed in terms of the variables $x_1$ and $r_Y$:
\begin{subequations}
\begin{eqnarray}
E_2^\ast &=& \frac{E_b}{2 r_X} (r_X^2-r_Y^2+r_c), \\
\label{absp1}%
|\bm{p}_1| &=&E_b (x_1^2-r_c)^{1/2},
\\
\label{absp2}%
|\bm{p}_2^*| &=&\frac{E_b}{2 r_X}
\lambda^{\frac{1}{2}}(r_X^2, r_Y^2, r_c),
\\
\label{absp3}%
|\bm{p}_3^*| &=&\frac{r_Y}{2}E_b,
\end{eqnarray}
\label{absp}%
\end{subequations}
where $r_X=m_X/E_b$.
Substituting Eqs.~(\ref{xi-def}) and (\ref{absp}) into Eq.~(\ref{dphi4}),
we obtain
the four-body phase space $d\Phi_4$ in terms of $x_1$ and $r_Y$:
\begin{equation}
d\Phi_4
=
\frac{E_b^4}{2^{12} \pi^7}
\frac{r_Y (x_1^2-r_c)^{1/2} \lambda^{1/2}(r_X^2, r_Y^2, r_c)}{r_X^2}
dx_1 dr_Y d\Omega_2^* d\Omega_3^*.
\label{PS4}%
\end{equation}

\end{document}